\documentclass[amsmath,amssymb,11pt]{article}
\usepackage{jheppub2}
\pdfoutput=1

\usepackage{amsmath,amssymb,amsthm,mathrsfs,bbm,bm}
\usepackage{latexsym,amscd,amsbsy,amsfonts,dsfont}
\usepackage{multirow}
\usepackage{graphicx}
\usepackage{xcolor}
\usepackage{caption}
\usepackage{comment} 
\usepackage{wrapfig}

\def\la{\label}

\usepackage{changepage}

	\newcommand{\lp}{\left (}
	\newcommand{\rp}{\right )}

\newcommand{\Glo}{\Gamma_{\text{\tiny LO}}}

\newcommand*{\affmark}[1][*]{\textsuperscript{#1}} 

\catcode`,\active

\catcode`\,12

\numberwithin{equation}{section}  
\allowdisplaybreaks 

\title{Not all black holes decohere quantum superpositions}

\author{Anna Biggs\affmark[1],}
\emailAdd{abiggs@princeton.edu}
\author{Stefano Trezzi\affmark[2]}
\emailAdd{strezzi@icc.ub.edu}

\affiliation{
\affmark[1]Jadwin Hall, Princeton University, Princeton, NJ 08540, USA\\
}

\affiliation{
\affmark[2]Departament de Física Quàntica i Astrofísica and
  Institut de Ciències del Cosmos,
 Universitat de Barcelona, 08028 Barcelona, Spain\\
}

\abstract{We study the decoherence induced by near-extremal charged black holes on quantum systems in their exterior. Specifically, we analyze a thought experiment recently discussed in the literature, where the quantum system is a charged particle prepared in a spatial superposition. Near-extremal black holes are known to exhibit large quantum metric fluctuations of the near-horizon geometry at low temperatures. We show that, at late times, if the black hole is sufficiently close to extremality, these quantum gravity effects make the decoherence rate vanish. This phenomenon is due to a spin-induced energy gap in the quantum black hole spectrum. For energies above the gap, the decoherence rate becomes nonzero, but is still suppressed relative to semiclassical expectations, so these quantum gravity effects always enhance the coherence of the superposition.}

\begin{document}

\maketitle

\section{Introduction}
\label{sec:intro}

A central hypothesis in the study of quantum black holes is the idea that, from the outside, a black hole is described by a unitary quantum system with a number of internal degrees of freedom of order the black hole entropy. One strategy for learning about these degrees of freedom is to study them operationally, by comparing the observable properties of black holes to those of more familiar quantum systems. An interesting observable to consider is the decoherence that a black hole induces on its environment. 

When coupled to a qubit via the electromagnetic or gravitational field, finite-temperature  matter is expected to decohere the qubit at a constant rate at late times. 
The rate itself depends on the resistivity or viscosity of the material \cite{Biggs:2024dgp}. This effect can be understood as the result of the qubit becoming entangled with the microscopic matter degrees of freedom.

Recent work has shown that black holes also produce this effect in semiclassical gravity \cite{Danielson:2022tdw,Danielson:2022sga,Gralla:2023oya}.\footnote{When the qubit is sourcing an electromagnetic field, the rate of decoherence induced by the black hole is comparable to that of ordinary matter. In other words, it is easy to find ordinary matter with the same ``resistivity'' as the black hole. In the gravitational case, the effective black hole ``viscosity'' is high compared to most matter systems.} Further exploration of this effect has been carried out in \cite{Wilson-Gerow:2024ljx,Danielson:2024yru,Li:2024guo,Li:2024lfv,Danielson:2025iud,Kudler-Flam:2025yur,Danielson:2025aji}. From the semiclassical metric description, the existence of a Hilbert space of black hole microstates which become  entangled with the qubit is not manifest. However, their role in the decoherence effect can be inferred indirectly from the causal structure of the horizon. In this respect, the decoherence effect can be viewed as illustrating another way in which semiclassical gravity appears to ``know'' about its own microstates. From the standpoint of QFT on the curved black hole background, we would say that the qubit becomes entangled with the black hole by virtue of the field it sources falling behind the horizon. 
In other words,  the black hole degrees of freedom responsible for the decoherence have an emergent semiclassical description in terms of a spacetime with a horizon.  So, the finding that typical black holes decohere their surroundings once more reinforces the view that black holes admit a conventional quantum Hilbert space description.

On a separate front, recent insights on near-extremal black holes have shown that, at very low temperatures, these black holes exhibit large quantum metric fluctuations of the near-horizon geometry which significantly modify observables such as their absorption cross-section and emission spectrum \cite{Brown:2024ajk,Mohan:2024rtn,Maulik:2025hax,Emparan:2025sao,Biggs:2025nzs,Lin:2025wof,Emparan:2025qqf,Betzios:2025sct,Kraus:2025efu,Rakic:2025svg,Luo:2026epp}. For example, the low-frequency electromagnetic and gravitational absorption cross-section of near-extremal charged black holes is highly suppressed relative to the semiclassical prediction, which is due to a spectral gap in the quantum-corrected density of states \cite{Emparan:2025qqf}. A natural question is then how these quantum gravity effects modify the decoherence rate that near-extremal black holes induce on quantum systems in their exterior. We address this question in this work, specifically for near-extremal Reissner-Nordström black holes in four spacetime dimensions.

For concreteness, we analyze the thought experiment proposed in \cite{Danielson:2022sga,Danielson:2022tdw}, in the case where the quantum system being decohered is a simple spatial superposition of a single electrically charged particle. We are interested in the limit where the superposition---prepared by an experimenter typically called ``Alice''---is held stationary outside the black hole for a long proper time $T$, and any radiation emitted during the separation or recombination of the particle is made negligible.\footnote{This can always be obtained by performing the separation and recombination processes slowly enough.} The decoherence effect in this context has been  explained from a number of complementary perspectives, including as the result of a ``black hole memory effect'' wherein soft radiation emitted by the superposition is absorbed by the black hole, causing a change in the gauge potential on the horizon. See \cite{Satishchandran:2025cfk} for a recent review. Arguably the simplest and most general way to analyze this problem, and the one we will employ in this paper, is to derive a low-energy effective theory for the interaction between the black hole and the quantum superposition which only involves operators describing the electromagnetic multipole moments of the two systems. From this perspective, the decoherence arises due to thermal or quantum fluctuations of the black hole multipole operators and can be expressed purely in terms of their correlation functions.

These correlation functions capture the dynamics of fields which probe the near-horizon region of the black hole. For near-extremal black holes, the semiclassical description of the near-horizon region breaks down when the black hole energy above extremality, $E$, approaches the ``breakdown scale''
\begin{equation}
\label{eq:Eb_def}
    E_b = \frac{G_N}{r_0^3} = \frac{\pi}{r_0 S_0} \, ,
\end{equation}
which is the energy scale at which quantum gravity effects become important \cite{Iliesiu:2020qvm}. Here, $S_0$ and $r_0$ denote respectively the (naive) semiclassical entropy and horizon radius at extremality,\footnote{Viewed from the far zone, at leading order the black hole cannot be distinguished from a classical extremal solution characterized only by the scale $r_0$. 
The much lower temperature scale $1/\beta$ appears only in subleading corrections in the expansion away from the extremal geometry, controlled by $r_0/\beta$.} and $G_N$ is Newton's constant.

As the black hole approaches extremality, the classical near-horizon geometry becomes approximately $\text{AdS}_2 \times S^2$, where the length of the $\text{AdS}_2$ ``throat'' is controlled by the inverse Hawking temperature $\beta$. In the quantum regime $E \lesssim E_b$, the length of this throat exhibits a large quantum variance. In order to correctly describe the black hole spectrum and dynamics, this gravitational mode of the near-horizon geometry needs to be quantized explicitly. Remarkably, the fluctuations of this mode are large for macroscopic black holes, meaning that quantum gravity effects can be dominant in spacetime regions where the curvature is much smaller than the Planck scale. Moreover, this regime is accessible by gravitational path integral methods, and the mode in question is described by a quantum mechanical theory whose action is given by a Schwarzian time derivative \cite{Moitra:2019bub,Iliesiu:2020qvm,Iliesiu:2022onk,Kapec:2023ruw,Rakic:2023vhv}.
Correlation functions of the black hole multipole operators have been computed including the full quantum backreaction of the metric, via performing the gravitational path integral over this ``Schwarzian'' mode.  Here, we leverage those results to find the quantum gravity corrected decoherence rates.\footnote{Note that previous work analyzing the decoherence caused by extremal black holes \cite{Gralla:2023oya,Li:2024guo} has treated the spacetime semiclassically, but this treatment is incorrect below the energy scale \eqref{eq:Eb_def}.}

Previous work has considered Schwarzian corrections to decoherence in the case where the mediating field is a minimally coupled massless scalar field \cite{Li:2025vcm}. As we currently lack evidence for a fundamental long-range scalar interaction in nature, here we consider decoherence due to interactions mediated by the electromagnetic field. We find that when the black hole's energy above extremality is below the scale  \eqref{eq:Eb_def}, the decoherence rate that it induces on Alice's superposition vanishes in the large $T$ limit. We demonstrate this explicitly at leading order and next-to-leading order in the coupling between the black hole and Alice's system, and conjecture this to be true at any order. Above the threshold \eqref{eq:Eb_def}, the decoherence rate is nonzero but suppressed relative to the semiclassical prediction, meaning that Schwarzian quantum gravity effects enhance the coherence of the quantum system.  In short, the decoherence effect observed in \cite{Danielson:2022tdw,Danielson:2022sga} is not a universal feature of black holes in quantum gravity.

Our result is closely related to the fact that such black holes are transparent, at leading order, to low-frequency electromagnetic radiation at energies below $E_b$ 
\cite{Emparan:2025qqf, Lin:2025wof}. In particular, there is a spin-induced gap in the microscopic density of states (cf.~\eqref{eq:gap}) which is responsible for both effects. However, these observables differ when we look to higher orders in the effective coupling between the black hole and qubit system. Multi-photon processes with zero total angular momentum allow the absorption cross-section to be nonzero, while such processes do not contribute to the decoherence. The decoherence is therefore an independent probe of the quantum metric fluctuations of the near-horizon geometry.

The aforementioned energy gap in the quantum density of states is given  by \cite{Iliesiu:2020qvm}
\begin{equation}
\label{eq:gap}
    E_{0,j} = \frac{j \, (j + 1)}{2} \, E_b \, ,
\end{equation}
where $j$ is the spin of the black hole as measured in units of $\hbar$. Here $E_{0,j} \equiv M_{0,j} - M_0$ denotes the energy gap above the spinless extremal state, where $M_0 = r_0/G_N = Q / \sqrt{G_N}$ is the mass of the extremal static black hole, and $M_{0,j}$ is the mass of the lowest-energy state for a black hole with spin $j$. In particular, if our initially non-rotating black hole starts with $j=0$, by absorbing or emitting a spin-$1$ photon (p-wave) it will transition to a $j=1$ state, which is sensitive to the gap \eqref{eq:gap}. In particular, in the $j=1$ case the energy gap \eqref{eq:gap} is simply equal to the breakdown scale \eqref{eq:Eb_def}, $E_{0,j=1} = E_b$. This is why there is a fundamental qualitative difference between the (spinless) scalar case considered in \cite{Li:2025vcm} and the electromagnetic case that we are considering. In the scalar case this energy gap is absent, so there is no transparency effect, and the decoherence rate is always nonzero.

Our finding that the decoherence rate of near-extremal black holes vanishes at late times is again consistent with what one would expect for an ordinary quantum system. At low temperatures, quantum effects typically become important, often yielding selection rules for the interaction between the quantum system and its environment. If those selection rules forbid a certain interaction channel, that channel of the environment cannot decohere the system. In this work we show that a similar phenomenon occurs for black holes, with the selection rules being controlled by the spin-induced energy gap \eqref{eq:gap}.

\paragraph{Gravitational case.}  In this work we restrict our attention to the electromagnetic case, but we expect similar qualitative conclusions to hold for decoherence mediated by the gravitational field. 
This is because gravitons are spin-$2$ particles, and like photons their absorption is subject to selection rules controlled by the spectral gap \eqref{eq:gap}. However, there will be some technical differences from our analysis, because in the gravitational case the effective description we will adopt in Section~\ref{sec:setup} must be modified to take into account the electromagnetic-gravitational mixing between $\ell \ge 2$ photon and graviton modes in the Reissner–Nordström background \cite{Crispino:2009zza,Oliveira:2011zz}.

\paragraph{Generalizations.} While our calculations refer to charged black holes in 4D Einstein-Maxwell theory, our results extend naturally to black holes carrying different conserved charges, and to higher dimensional scenarios. In particular, we expect a similar result for any conserved charge whose quantization leads to gaps in the black hole spectrum \cite{Emparan:2025qqf}. The case of near-extremal Kerr requires extra care: technically, due to the lack of spherical symmetry, and physically due to superradiant emission, which drives the black hole away from extremality.\footnote{As argued in \cite{Kapec:2023ruw,Rakic:2023vhv}, rotating black holes also exhibit large quantum fluctuations of the near-horizon geometry when the energy above extremality is of order \eqref{eq:Eb_def}. Superradiant emission, however, precludes the possibility of maintaining such low temperatures over a timescale $\gtrsim E_b^{-1}$, as would be required to observe these quantum effects. Even setting superradiance aside, the smallness of the quantum scale \eqref{eq:Eb_def} further prevents the possibility of observation for astrophysical black holes. For a solar-mass black hole, this scale corresponds to a temperature of order $10^{-82}$~K, or equivalently a thermal wavelength $\sim 10^{79}$~m.} We expect similar results to those we find here for near-BPS black holes, because their spectrum is also gapped. Note that this gap is of a different nature than the spin-induced gap \eqref{eq:gap} in the non-supersymmetric case, since it is also present for the spin-zero states in the supermultiplet and may be regarded as an enhanced level repulsion of the highly degenerate BPS states \cite{Emparan:2025qqf}. We elaborate more on this in Section~\ref{sec:conclusions}.

\paragraph{Outline.} The rest of the paper is organized as follows. 
In Section~\ref{sec:setup} we introduce the setup in more detail and review the effective theory description of the coupling between the black hole and Alice's system. In Section~\ref{sec:time_evol_RDM} we compute the time evolution of Alice's reduced density matrix and we relate it to the decoherence rate. We then perform the computation of the decoherence rate in the quantum Schwarzian theory in Section~\ref{sec:Gamma_Sch}. We briefly conclude  in Section~\ref{sec:conclusions} and mention possible future directions. The appendices contain various calculations underlying Sections~\ref{sec:time_evol_RDM} and \ref{sec:Gamma_Sch}.

\section{Setup and effective theory approach}
\label{sec:setup}

We briefly review the setup of the decoherence thought experiment discussed in \cite{Danielson:2022sga, Danielson:2022tdw, Gralla:2023oya}. 

An experimenter, Alice, prepares a spatial quantum superposition in the exterior of a black hole. Here we specifically consider a near-extremal Reissner-Nordström black hole in 4D. Following \cite{Biggs:2024dgp}, we take Alice's system to be an electric dipole $\vec{P}_{A}$ prepared in a superposition of two opposite orientations.\footnote{More explicitly, we can consider a particle with electric charge $q$ in a superposition of localized position-space wavepackets $|\Psi\rangle = \frac{1}{\sqrt{2}}\lp |-\frac{d}{2}\rangle + |+ \frac{d}{2}\rangle \rp $, where $d$ is the distance between the spatial positions. Between these two positions is an oppositely charged particle $-q$ which remains stationary throughout the experiment. Then, $|\vec{P}_{A}| = q d/2$.} After the superposition is prepared, it remains stationary outside the black hole for a long proper time $T$. The dipole is turned on and off slowly, over a time $T_\text{turn} \gg |\vec{P}_{A}|$ (but still small compared to $T$,  i.e. $T \gg T_\text{turn}$), so that radiation to infinity is negligible. Denoting Alice's radial position by $\vec{b}$, we are interested in the late time limit where $T \gg \beta \gg |\vec{b}| \gg r_0$, where $r_0$ is the horizon radius and $\beta$ is the black hole inverse temperature.\footnote{We will later work in the microcanonical ensemble and label black hole states by their energy $E \equiv M - M_0$ above extremality. We will then employ the relation \eqref{eq:beta_E} to translate between energy $E$ and inverse temperature $\beta$ near extremality.}

In this limit, the black hole can be approximated by a point particle in flat space, and its interaction with Alice's quantum system is dominated by the coupling between their electromagnetic dipole moments. We will briefly explain this effective description, leaving details to \cite{Biggs:2024dgp}. 

\begin{figure}[t!]
    \centering
    \includegraphics[width=0.3 \textwidth]{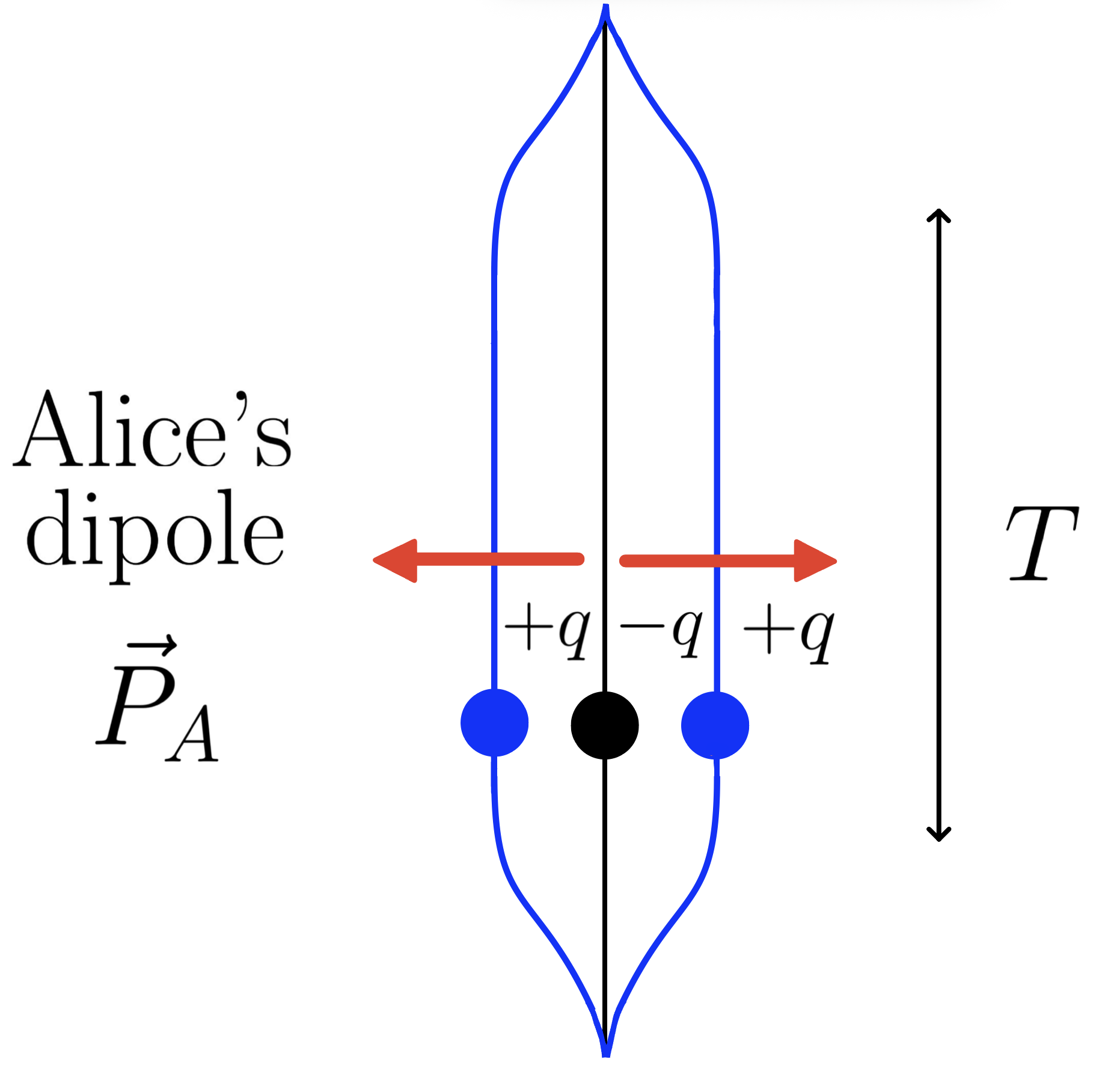}\qquad\qquad\includegraphics[width=0.6 \textwidth]{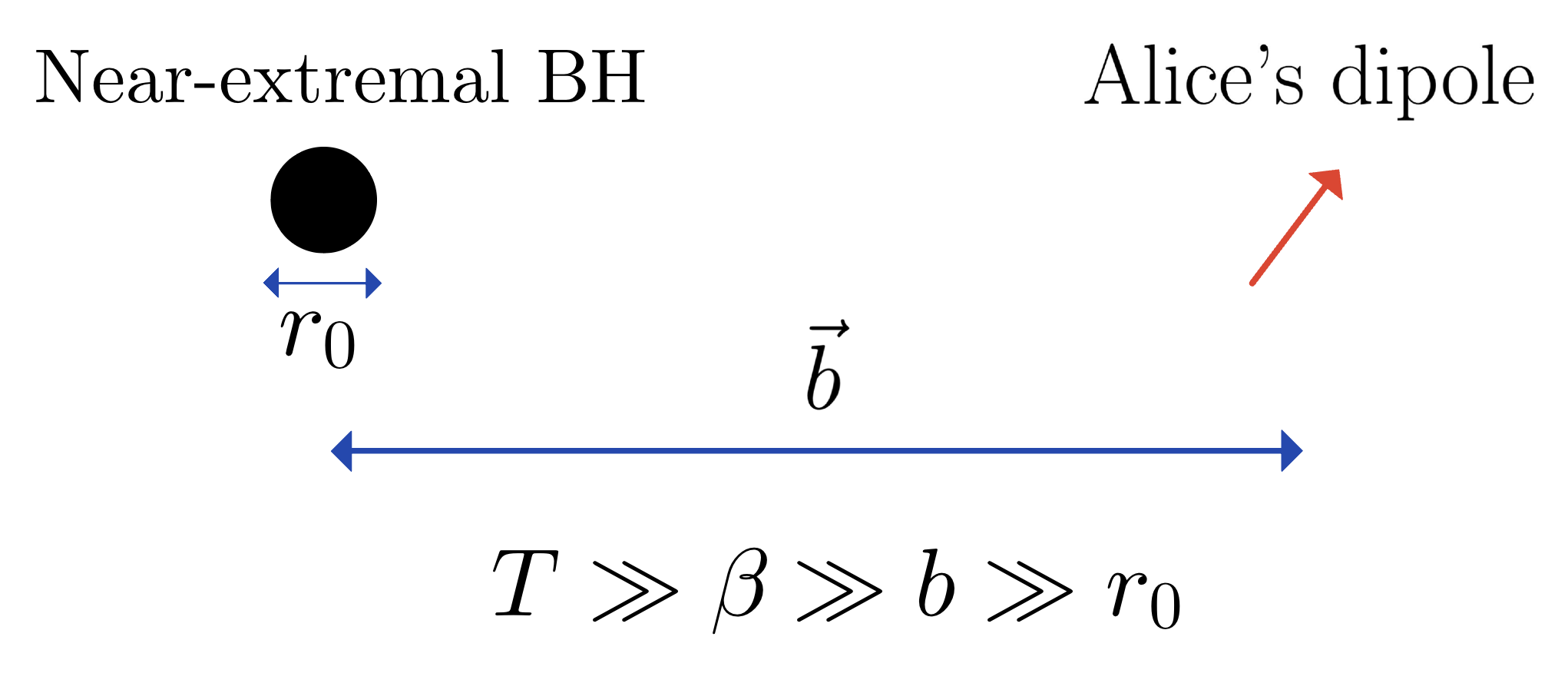}
    \caption{\small Illustration of the quantum system in the black hole exterior (left) and the relevant scales in the problem (right). Alice's state is  the superposition of an electric dipole pointing in two opposite directions. We consider a low-energy, long-time limit where the black hole can be treated as a point particle in flat space.}
    \label{fig:decoherence_setup}
\end{figure}

The electromagnetic field sourced by $\vec{P}_{A}$ has a wavelength $1/\omega \sim T \gg r_0$, so it is not sensitive to the finite size of the black hole. To leading order in $r_0 \omega$,  interactions of the electromagnetic field with the black hole are captured by the coupling to its electric dipole moment $\vec{P}_{B}$, which is an operator living on the effective point particle worldline of the black hole.  The total action in this effective description is
\begin{equation}
\label{eq:full_I}
    I_\text{tot} = I_\text{Sch} + \frac{1}{2 e^2} \, \int d^4 x \, (\vec{E}^2 - \vec{B}^2) - \int dt \, \vec{P}_B(t) \cdot \vec{E}(t) - \int dt \, \vec{P}_A(t) \cdot \vec{E}(t) \, \sigma_3 \, .
\end{equation}
Here $\vec{P}_{B}$ acts on the black hole Hilbert space $\mathcal{H}_B$ and $\sigma_3$ on Alice's quantum system $\mathcal{H}_A$; the total Hilbert space is $\mathcal{H} = \mathcal{H}_A \otimes \mathcal{H}_B$. The eigenvalues of the third Pauli matrix $\sigma_3 $ parameterize the two opposite orientations of Alice's dipole moment, and $\vec{P}_A$ is a classical vector indicating its direction.  $I_\text{Sch}$ is the Schwarzian action that describes the black hole in isolation.

Note that, in the vicinity of a charged black hole, modes of the gravitational field mix with $\ell \geq 2$ modes of the electromagnetic field, where $\ell$ is the angular momentum of the field perturbation $\phi$ in the spherical harmonics decomposition $\phi(t,r,\theta,\varphi) = e^{-i \omega t} \, u(r) \, Y_{\ell m}(\theta,\varphi)$ on the black hole background. This means that the low-energy effective description \eqref{eq:full_I} is accurate only for the lowest energy perturbations that correspond to exciting the $\ell = 1$ mode of the electromagnetic field. Since our interest is the dipole-dipole interaction at low frequencies, this action is sufficient for our purposes. 

$\vec{P}_A$ varies over a timescale $T \gg |\vec{b}|$ and $\vec{P}_B$ over a timescale $\beta \gg |\vec{b}|$. In this limit, we can integrate out the mediating field completely,  leaving the following dipole-dipole interaction:
\begin{equation}
\label{eq:I_int}
    I_\text{int} = \int dt \, \sigma_3 \, O(t) 
    \, , \qquad H_\text{int}(t) = -\sigma_3 \, O(t) \, ,
\end{equation}
where
\begin{equation}
\label{eq:Odef}
    O(t) = -\frac{e^2}{4 \pi b^3} \, (\delta_{ij} - 3 \hat{b}_i \hat{b}_j)  P_A^i \, P_B^j(t) ~~,~~~~~~\hat{b} = \frac{\vec{b}}{|\vec{b}|} \, .
\end{equation}

In subsequent sections, we will express the decoherence in terms of correlation functions of the electromagnetic dipole operator $P^{i}_B(t)$. We then evaluate these correlation functions in the Schwarzian theory, which captures the effect of the near-horizon quantum fluctuations on the black hole's response to external perturbations.

\section{Time evolution of Alice's reduced density matrix}
\label{sec:time_evol_RDM}

We start with Alice's system in a generic initial state described by the $2 \times 2$ density matrix\footnote{The specific density matrix corresponding to the setup described in Section~\ref{sec:setup} is the pure state 
\begin{equation*}
    \rho_A = \frac{1}{2} \,
    \begin{pmatrix}
        1 & 1\\
        1 & 1
    \end{pmatrix}
    \, .
\end{equation*}
}
\begin{equation}
\label{eq:rhoA_general}
    \rho_A =
    \begin{pmatrix}
        \rho_{++} & \rho_{+-}\\
        \rho_{+-}^\ast & \rho_{--}
    \end{pmatrix}
    \, .
\end{equation}
The black hole begins in a stationary state $\rho_B$, which we will later take to be a fixed-energy state in the microcanonical ensemble. Let the initial state of the total system be a product state, $\rho(t=0) = \rho_A \otimes \rho_B$. Then the state of Alice's system at time $T$, $\rho_A(T)$, is given by
\begin{equation}
\label{eq:rhoAT}
    \rho_A(T) = \text{tr}_B\!\left[\mathcal{T} \exp\!\left(-i \, \int_0^T dt \, H(t) \right) \,\, (\rho_A \otimes \rho_B) \,\, \overline{\mathcal{T}} \exp\!\left(i \, \int_0^T dt \, H(t) \right) \right] \, ,
\end{equation}
where $\mathcal{T}$ denotes the time-ordering operator and $\overline{\mathcal{T}}$ the anti-time-ordering operator. Considering now an effective action of the form \eqref{eq:I_int}, the time evolution of the off-diagonal components $\rho_{+-}$ and $\rho_{-+} = \rho_{+-}^\ast$ of Alice's reduced density matrix \eqref{eq:rhoAT} can be expressed as
\begin{align}
\label{eq:F_def}
    \rho_{+-}(T) = F(T) \, \rho_{+-} \, ,
\end{align}
where we have defined the influence functional
\begin{equation}
\label{eq:FTdef}
   F(T) = \text{tr}_{B}\!\left[ \mathcal{T} \exp\!\left(i \, \int_0^T dt \, O(t)\right) \, \rho_B \, \overline{\mathcal{T}} \exp\!\left(i \, \int_0^T dt \, O(t)\right) \right] \, .
\end{equation}
Equivalently, we can define the cumulant-generating functional $\Phi(T)$, also known as the decoherence function, the expansion of which contains only connected correlators \cite{Breuer}:
\begin{equation}
    \rho_{+-}(T) = e^{-\Phi(T)} \, \rho_{+-} \, , \qquad \Phi(T) \equiv -\log F(T) \, .
\end{equation}
On the other hand, the diagonal elements of $\rho_A$ do not evolve in time. The resulting dynamics is therefore a pure dephasing (i.e., ``phase-damping'') channel for Alice’s qubit.

Expanding $F(T)$ to fourth order in $O(t)$, we find 
\begin{samepage}
\begin{align}
    \label{eq:F0_F2}
    F(T) =~& 1  - 2 \, \int_0^T dt_1 \int_0^T dt_2 \, \langle O(t_1) O(t_2) \rangle\\
    \label{eq:F4_1}
    &+\frac{1}{12} \, \int_0^T dt_1 \int_0^T dt_2 \int_0^T dt_3 \int_0^T dt_4 \, \Big(\langle\mathcal{T}\{O(t_1) O(t_2) O(t_3) O(t_4) \} \rangle + \text{c.c.} \Big)\\
    \label{eq:F4_2}
    &+\frac{1}{2} \, \int_0^T dt_1 \int_0^T dt_2 \int_0^T dt_3 \int_0^T dt_4 \, \langle \overline{\mathcal{T}}\{O(t_1) O(t_2)\} \, \mathcal{T}\{O(t_3) O(t_4)\} \rangle + \cdots  \, ,
\end{align}
\end{samepage}
where the complex conjugate ``c.c.'' refers to the completely anti-time-ordered correlator $\langle\overline{\mathcal{T}}\{O(t_1) O(t_2) O(t_3) O(t_4) \} \rangle$. Here we assumed that all odd-point functions of $O$ vanish. This is true in our setting because the electric dipole operator $\vec{P}_B$ is odd under spatial parity, while the initial state of the black hole $\rho_B$ is parity-invariant. Exploiting the series $\log(1 + x) = x - \frac{x^2}{2} + O(x^3)$, we can expand $\Phi(T)$ in a similar way to \eqref{eq:F0_F2}--\eqref{eq:F4_2}, but now the expansion will only contain connected correlators of $O$.

We now use time translation invariance of the correlators, as well as the assumption that the correlators decay in time separation after a timescale of order $\beta \ll T$, so that $\Phi(T)$ becomes linear in $T$ in the large $T$ limit:\footnote{Note that our definition of the decoherence rate and the one adopted in \cite{Biggs:2024dgp,Li:2025vcm} differ by a factor of two, $\Gamma_\text{ours} = 2 \, \Gamma_\text{theirs}$.}
\begin{equation}
\label{eq:Phieqn2}
    \Phi(T) \sim \Gamma \, T \, , \qquad \Gamma = \Gamma_{\text{\tiny LO}} + \Gamma_{\text{\tiny NLO}} + \cdots \, ,
\end{equation}
where $\Gamma$ is called the decoherence rate, and $\Gamma_{\text{\tiny LO}}$ and $\Gamma_{\text{\tiny NLO}}$ denote its leading order and next-to-leading order in the interaction \eqref{eq:I_int}. The leading order contribution is given by 
\begin{equation}
\label{eq:GammaLO}
     \Gamma_{\text{\tiny LO}} =  2 \, \int_{-\infty}^{+\infty} dt \, \langle O(t) O(0) \rangle \, ,
\end{equation}
while the explicit expression for the next-to-leading order contribution will be given below (cf.~\eqref{eq:GammaNLO}).

Note that in doing perturbation theory we have assumed that \eqref{eq:I_int} is weakly coupled. One way of seeing that this is the case is by calculating the transmission probability of low-frequency electromagnetic waves through the gravitational potential barrier that separates the near-horizon $\text{AdS}_2$ region from the asymptotically flat region. In the limit $\omega \ll 1/\beta$, the transmission probability $P_\text{gb}$ (also known as the greybody factor) scales as \cite{Bai:2023hpd, Brown:2024ajk}
\begin{equation}
    P_\text{gb} \propto (r_0 \omega)^4 \left(\frac{r_0}{\beta} \right)^4 \, ,
\end{equation}
which is small for the low-frequency $r_0 \omega \ll 1$ and near-extremal $r_0/\beta \ll 1$ limits that we consider here.

\section{Decoherence rate in the quantum Schwarzian theory}
\label{sec:Gamma_Sch}

The correlators in \eqref{eq:F0_F2}--\eqref{eq:F4_2}, which describe the black hole response to external electromagnetic perturbations, have been computed previously
including the effect of quantum gravity fluctuations in the $\text{AdS}_2$ throat. See \cite{Mertens:2022irh} for a recent review. Making use of those results requires a minor translation between our $O(t)$, defined in \eqref{eq:Odef}, and the operators that couple to the Schwarzian theory in their conventional normalization.

In the near-horizon region, modes of the coupled electromagnetic and gravitational fields behave as independent 
degrees of freedom with an effective mass set by $\ell$, the angular momentum quantum number. This determines the conformal dimension $\Delta$ of the corresponding operator in the dual quantum mechanical theory. In particular, the field perturbation acts as a source at the ``mouth of the throat'' (namely, the boundary of the $\text{AdS}_2$ region), and we will denote the corresponding response operator of conformal dimension $\Delta$ by $\mathcal{O}_\Delta^{(\ell,m)}(t)$, where $m$ is the azimuthal mode number carried by the perturbation in its spherical harmonics decomposition. For photons, the lowest physical mode is the dipole, $\ell = 1$, which corresponds to an operator with conformal dimension $\Delta = 3$. Converting $P_B^{i}$ from the Cartesian to the spherical basis via the spherical-tensor projection $\hat{e}_{\ell, m}^i$,
\begin{equation}
\label{eq:PBi_PB1m}
    P_B^i(t) = \sum_m \hat{e}_{1, m}^i \, P^B_{1 m}(t) \, ,
\end{equation}
where
\begin{equation}
\label{eq:hat_e}
    \hat{e}_{1, \pm 1} = \mp \frac{\hat{x} \pm i \hat{y}}{\sqrt{2}} \, , \qquad \hat{e}_{1, 0} = \hat{z} \, ,
\end{equation}
then $P_{1m}^B$ is equivalent to the  response operator $\mathcal{O}_3^{(1,m)}$ up to a dimensionful proportionality constant. For simplicity, we consider the case where $\vec{P}_A$ points radially in the $\hat z$ direction, $\vec{P}_A \propto \hat{b}$ with $\hat{b} = \hat{z}$.  Then, only the $m = 0$ component of $\mathcal{O}_{3}^{(1,m)}$ appears in the correlators. In this case, \eqref{eq:GammaLO} becomes
\begin{align}
    \label{eq:Gamma_LO_sO}
    \Gamma_{\text{\tiny LO}} &= 2 \, \mathcal{N}^2 \, \int_{-\infty}^{+\infty} dt \, \langle \mathcal{O}(t) \mathcal{O}(0) \rangle \, 
\end{align}
where for notational simplicity we have suppressed the operator subscripts, $\mathcal{O}(t) \equiv \mathcal{O}_3^{(1,0)}(t)$, and $\mathcal{N}$ is the proportionality constant between $O(t)$ and $\mathcal{O}(t)$,
\begin{equation}
    O(t) = \mathcal{N} \, \mathcal{O}(t) \, ,
\end{equation}
which scales as 
\begin{align}
\label{eq:sN}
    \mathcal{N} \propto \frac{e q d r_0^4}{b^3} \, .
\end{align}
Eq.~\eqref{eq:sN} can be derived by requiring that the quantum-corrected decoherence rate reduces to the semiclassical one in the limit where the black hole energy is much larger than the quantum scale \eqref{eq:Eb_def}. For details on this matching procedure, see Appendix~\ref{matching}.

Since the precise value of the constant \eqref{eq:sN} depends on the particular configuration being considered (namely, the relative orientation between Alice's dipole $\vec{P}_A$ and the radial vector $\vec{b}$ between Alice's lab and the black hole), in the following we will not report the numerical prefactors in front of our expressions. The precise prefactors corresponding to our particular choice of configuration $\vec{P}_A \propto \hat{b} = \hat{z}$ can be found in Appendix~\ref{matching}.

\subsection{Leading order}
\label{subsec:Gamma_LO}

In this section we evaluate the leading-order decoherence rate $\Gamma_{\text{\tiny LO}}$. It was shown in \cite{Emparan:2025qqf} that the low-frequency photon absorption cross-section is identically zero for black holes sufficiently near extremality.\footnote{This is true for coherent, circularly polarized waves. The black hole could absorb two-photon spin-singlet states from an unpolarized or linearly polarized wave, although this is a higher-order phenomenon and therefore strongly suppressed \cite{Brown:2024ajk, Emparan:2025qqf}.} This is because the black hole density of states vanishes when the black hole energy $E$ above extremality lies inside the spin-induced energy gap \eqref{eq:gap} in the microscopic density of states \cite{Heydeman:2020hhw}. Since the black hole cannot absorb single photons, one might expect the leading-order decoherence rate to vanish as well. We will see that this is indeed the case.

We will work in the microcanonical ensemble, where the energy $E$ above extremality, the spin $j$, and the azimuthal mode number $m_\text{\tiny BH}$ of the black hole are fixed.\footnote{In the quantum Schwarzian regime $E\lesssim E_b$, the condition $T \gg\beta$ implies $T \gtrsim E_b^{-1}$. A black hole that begins in the microcanonical ensemble will evolve through Hawking emission into a different ensemble in a time $\sim E_b^{-1}$ \cite{Biggs:2025nzs}.
This does not change the conclusion that coherence is preserved when $E < E_b$. It modifies our expressions for the decoherence rate (for $E > E_b$) by an $O(1)$ numerical prefactor which we will ignore here.} Inserting a resolution of the identity $\int dE' \, \rho_{j'}(E') \, |E', j', m'_\text{\tiny BH} \rangle \langle E', j', m'_\text{\tiny BH} | = \mathbb{I}$ and  performing the time integral in the two point function \eqref{eq:Gamma_LO_sO}, we have
\begin{align}
    \label{eq:Gamma_LO_micro_1}
    \Gamma_{\text{\tiny LO}} 
    &\propto \mathcal{N}^2 \, \rho_{j'}(E) \, \left|\langle E, j', m'_\text{\tiny BH} | \mathcal{O}(t=0) | E, j, m_\text{\tiny BH} \rangle \right|^2 \, ,
\end{align}
where $\rho_{j'}(E)$ is the density of states for a black hole with spin $j'$.

For simplicity, from now on we will restrict our attention to the case of a spinless black hole, $j=0$. As a consequence, we have that $m_\text{\tiny BH}=0$ and $j' = \ell = 1$. Moreover, since our configuration only involves the $m = 0$ component of the operator, also $m'_\text{\tiny BH}=0$.

The expressions for the JT gravity density of states and matter correlators have been derived in the literature on the Schwarzian theory, as reviewed in \cite{Mertens:2022irh}. The density of states of a generic black hole state $|E,j,m_\text{\tiny BH} \rangle$ has the following explicit expression:
\begin{equation}
    \label{eq:rho_j_E}
    \rho_{j}(E) = \frac{(2 j + 1) \, e^{S_0}}{2 \pi^2 E_b} \, \sinh\!\left(\! 2 \pi \, \sqrt{\frac{2 \, (E - E_{0,j})}{E_b}} \right) \, \theta(E - E_{0,j}) \, ,
\end{equation}
which vanishes inside the spin-induced energy gap \eqref{eq:gap} due to the Heaviside step function $\theta(E - E_{0,j})$. Here $S_0 = \pi r_0^2 / G_N$ is the naive semiclassical entropy of the black hole at extremality, while $E_b$ is the quantum scale \eqref{eq:Eb_def}.

The transition probability to go from the initial black hole state $|E,j=0,m_\text{\tiny BH} = 0 \rangle$ to a state $|E',j' = 1,m'_\text{\tiny BH} = 0 \rangle$ via the perturbation $\mathcal{O}(t) \equiv \mathcal{O}_3^{(1,0)}(t)$ is given by
\begin{equation}
    \label{eq:2pt_O}
    | \langle E', 1, 0 | \mathcal{O}(t=0) | E, 0, 0 \rangle |^2 = \frac{e^{-S_0} \, E_b^6}{11520} \, \Gamma\!\left(\!3 \pm i \sqrt{\frac{2 E}{E_b}} \pm i \sqrt{\frac{2 (E' - E_b)}{E_b}} \right) \, ,
\end{equation}
with the convention that the right-hand side is the product of the four Gamma functions that appear for each choice of sign.

\begin{figure}[t!]
    \centering
    \includegraphics[width=0.65\textwidth]{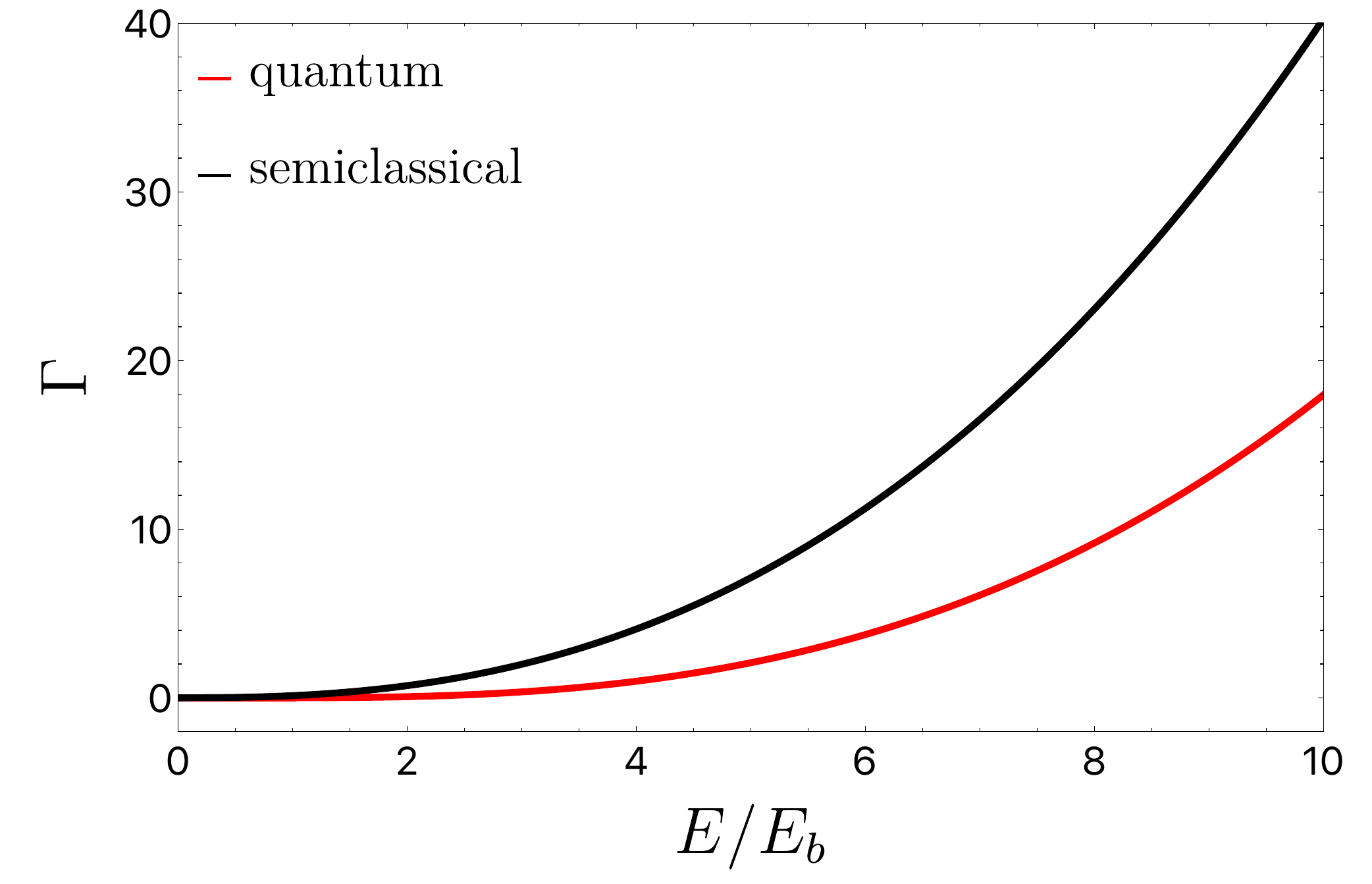}
    \caption{\small Comparison of the semiclassical \eqref{eq:Gamma_LO_sc_final_app} (black) and Schwarzian-corrected \eqref{eq:GammaLO_fully_explicit} (red) predictions for the electromagnetic decoherence rate $\Gamma$ as a function of the black hole energy above extremality $E/E_b$. On the vertical axis, the decoherence rate is expressed in units of $E_b^5 e^2 q^2 d^2 r_0^8 b^{-6}$ to make it dimensionless. Next-to-leading order effects will give small corrections to the two curves. The quantum gravity decoherence rate is always suppressed compared to the semiclassical one. The expressions adopted for this plot correspond to the specific configuration $\vec{P}_A \propto \hat{b} = \hat{z}$. Changing the orientation of $\vec{P}_A$  relative to $\hat b$ would affect the precise numerical values on the vertical axis but not the general behavior of the curves as a function of $E/E_{b}$}
    \label{fig:Ga_p_qusc}
\end{figure}

\begin{figure}[t!]
    \centering
    \includegraphics[width=0.475\textwidth]{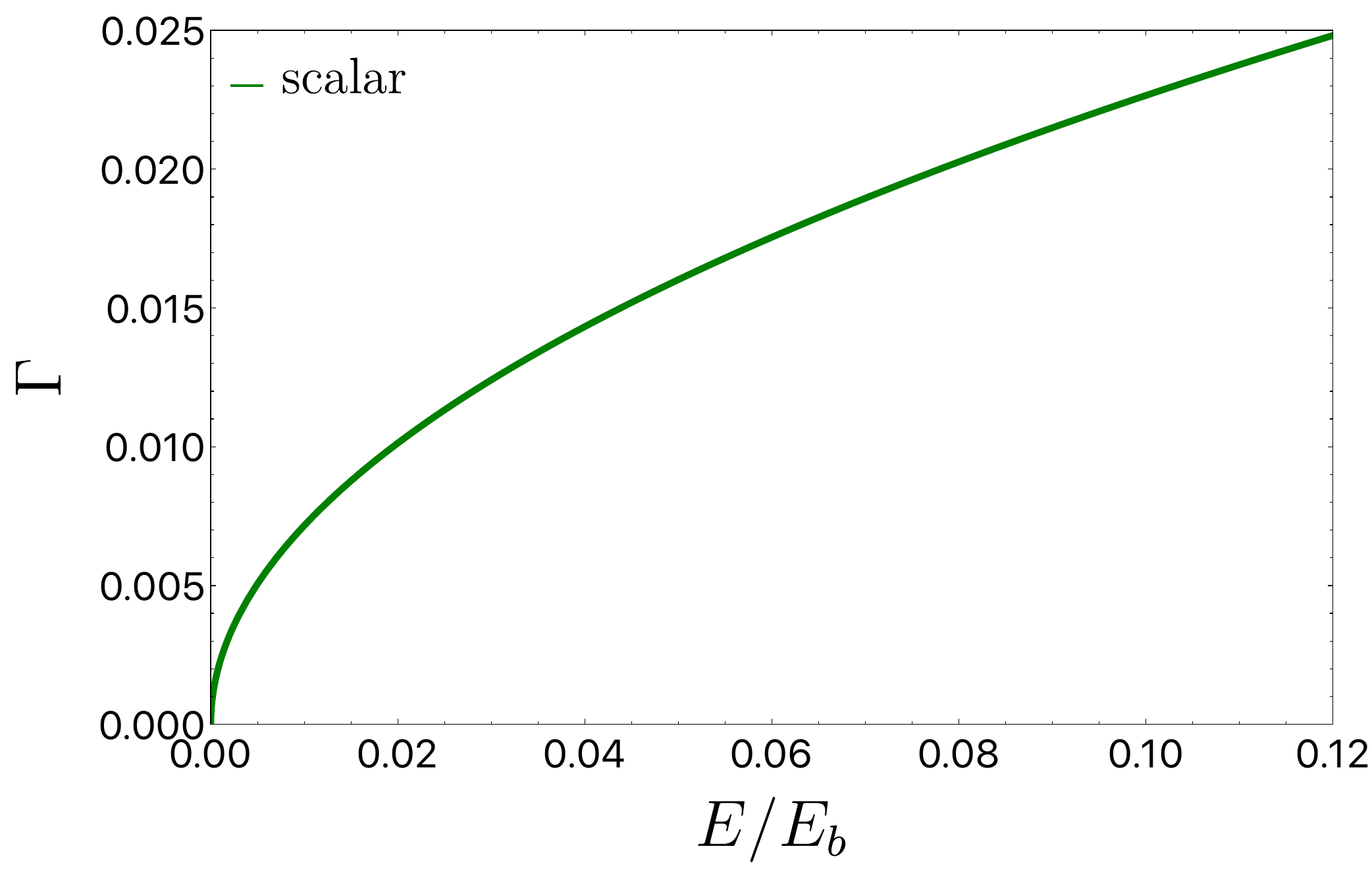}\qquad\includegraphics[width=0.475\textwidth]{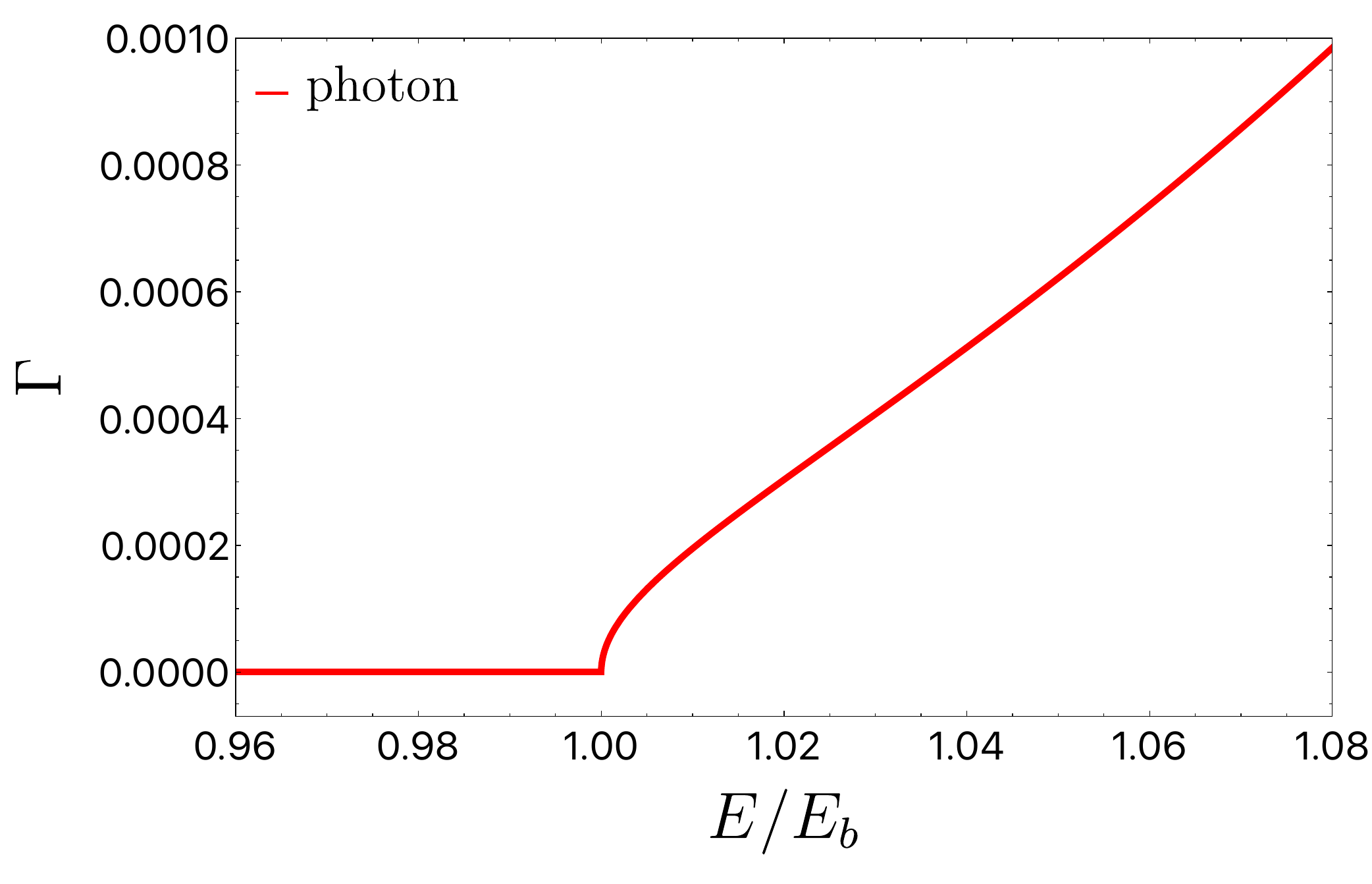}
    \caption{\small The Schwarzian-corrected decoherence rate $\Gamma$ as a function of $E/E_b$. We compare the case where the quantum system sources a scalar field \eqref{eq:Gamma_LO_scalar} (left) and an electromagnetic field \eqref{eq:GammaLO_fully_explicit} (right). The energy window \eqref{eq:GammaLO_zero_gap} where the decoherence rate vanishes extends up to $E/E_b = 1$ for the photon, while it is absent for the scalar. While higher order effects will give small modifications to the non-zero part of the curves, they do not modify the fact that $\Gamma = 0$ for $E < E_b$ in the electromagnetic case (cf.~\eqref{eq:GammaNLO_zero_gap} and \eqref{eq:Gamma_zero_gap}). The scalar and electromagnetic decoherence rates are expressed in units of $E_b j_A^2 r_0^2 b^{-2}$ and $E_b^5 e^2 q^2 d^2 r_0^8 b^{-6}$, respectively. As in Figure 2, one should not focus too much on the precise numerical values on the vertical axis, as these depend on the configuration of $\vec{P}_{A}$ relative to $\hat b$. 
}
    \label{fig:Ga_s_Ga_p}
\end{figure}

Comparing \eqref{eq:Gamma_LO_micro_1} with \eqref{eq:rho_j_E}--\eqref{eq:2pt_O}, we find
\begin{align}
    \Gamma_{\text{\tiny LO}} &\propto \mathcal{N}^2 \, \rho_1(E) \, | \langle E, 1, 0 | \mathcal{O}(t=0) | E, 0, 0 \rangle |^2\\
    &\propto \mathcal{N}^2 \, E_b^5 \, \sinh\!\left(\! 2 \pi \, \sqrt{\frac{2 \, (E - E_b)}{E_b}} \right) \, \Gamma\!\left(\!3 \pm i \sqrt{\frac{2 E}{E_b}} \pm i \sqrt{\frac{2 (E - E_b)}{E_b}} \right) \, \theta(E - E_b)\\
    \label{eq:Gamma_LO}
    &\propto \frac{E_b^5 e^2 q^2 d^2 r_0^8}{b^6} \, \frac{\left(8 \, \frac{E}{E_b} + 1 \right)^2 \, \sinh\!\left(2 \pi \sqrt{\frac{2 (E - E_b)}{E_b}} \right)}{\cosh\!\left(2 \pi \sqrt{\frac{2 E}{E_b}} \right) - \cosh\!\left(2 \pi \sqrt{\frac{2 (E - E_b)}{E_b}} \right)} \, \theta(E - E_b) \, .
\end{align}
Due to the step function $\theta(E - E_b)$ coming from $\rho_1(E)$, the leading-order decoherence rate is zero when $E < E_b$ (see the right panel of Figure~\ref{fig:Ga_s_Ga_p}):
\begin{equation}
\label{eq:GammaLO_zero_gap}
    \Gamma_{\text{\tiny LO}} = 0 \qquad \text{for } E < E_b \, .
\end{equation}

We emphasize that our statement applies to the contribution to the decoherence function $\Phi(T)$ which is linear in $T$ in the large $T$ limit. Such $T$ dependence is typical for environments which are approximately Markovian, or when the accumulated phase fluctuations induced by the environment have a finite correlation time and one is looking at timescales long compared to that correlation time. By ``decoherence rate'' we will always be referring to this linear-in-$T$ contribution to $\Phi(T)$. In general however, $\Phi(T)$ will contain additional contributions which are subleading in $T$ and may be nonzero even inside the gap $E < E_b$.

In the semiclassical regime $E \gg E_b$, \eqref{eq:Gamma_LO} reduces to
\begin{equation}
\label{eq:GaLOsc}
    \Gamma_{\text{\tiny LO}}^\text{sc} \propto \frac{e^2 q^2 d^2 r_0^8}{b^6} \, \beta^{-5} \, ,
\end{equation}
where the inverse Hawking temperature $\beta$ near extremality is related to the energy $E$ by
\begin{equation}
\label{eq:beta_E}
    \beta = \frac{2 \pi}{\sqrt{2 E_b E}} \, .
\end{equation}

In Figure~\ref{fig:Ga_p_qusc} we compare the Schwarzian-corrected and semiclassical results in the electromagnetic case ($\ell = 1$ photon perturbation). As in the case of the absorption cross-section \cite{Emparan:2025qqf}, the quantum curve always stays below the semiclassical one. As a consequence, even above the energy window \eqref{eq:GammaLO_zero_gap} where the decoherence rate vanishes, the coherence of the qubit system is enhanced by quantum gravity effects, as compared to the semiclassical expectation.

So far we have shown that the decoherence rate vanishes for $E < E_b$ at leading order in the coupling between the two systems. In Section~\ref{subsec:Gamma_NLO} we will argue that this is true at all orders in the coupling. The vanishing behavior of $\Gamma$ in energy window $E/E_b < 1$ is not easily distinguishable in Figure~\ref{fig:Ga_p_qusc}, but can be seen in Figure~\ref{fig:Ga_s_Ga_p}.

\paragraph{Connection to single-photon absorption.} The black hole absorption rate for a small time-dependent perturbation is given, at first order, by Fermi's golden rule (see eq.~(2.13) of \cite{Emparan:2025qqf}).\footnote{Here we will phrase our discussion in terms of the absorption rate, but we can equivalently rephrase it in terms of the emission rate since the two are related by $\omega \leftrightarrow -\omega$ by time-reversal invariance (cf.~\eqref{eq:Gamma_abs_em_sS}).} In particular, the zero-frequency limit of the absorption rate $\Gamma_\text{abs}(\omega)$ for an $\ell = 1$ photon perturbation is controlled by the same quantity that appears on the right-hand side of \eqref{eq:Gamma_LO_micro_1}, which corresponds to the expression of the leading-order decoherence rate $\Gamma_{\text{\tiny LO}}$ in the microcanonical ensemble.\footnote{We report a correction to eq.~(3.2) of \cite{Emparan:2025qqf}: the energies $E_i$ and $E_f$ must be shifted by the energy gaps $E_{0,j}$ and $E_{0,j'}$, respectively. Since we work in the $j=0$ sector, only the final energy $E_f$ shifts by $E_{0,\ell}$ (as $j'=\ell$). This does not affect the qualitative conclusions of \cite{Emparan:2025qqf}, but modifies the explicit expressions for quantum absorption cross-sections. A corrected version is forthcoming.} More precisely, we have the relation\footnote{At semiclassical level, in \cite{Wilson-Gerow:2024ljx} it was already observed that the decoherence rate is directly related to the low-frequency emission spectrum of the black hole, and locally in the frame of Alice's dipole, it can be viewed as a consequence of horizon radiation.}
\begin{equation}
\label{eq:GammaLO_Gammaabs}
    \Gamma_{\text{\tiny LO}} \propto \lim_{\omega \to 0} \frac{\Gamma_\text{abs}(\omega)}{\omega^3} \, .
\end{equation}
This means that the leading-order decoherence rate $\Gamma_{\text{\tiny LO}}$ vanishes precisely when the black hole is not able to absorb low-frequency $\ell = 1$ single-photon states. From \cite{Emparan:2025qqf}, we know that this happens when the black hole energy $E$ is inside the gap $E + \omega < E_b$, whose zero-frequency limit yields precisely the condition \eqref{eq:GammaLO_zero_gap}. We will elaborate more on this in Section~\ref{sec:rel_Green_abs}.

\paragraph{Comparison to scalar decoherence rate.} We compare our results for the $\ell = 1$ electromagnetic decoherence rate to those for an $\ell = 0$ minimally coupled, neutral massless scalar, which were obtained in \cite{Li:2025vcm}. 

In this case, the quantum decoherence rate in the microcanonical ensemble at leading order coincides with the semiclassical decoherence rate and is given (in our conventions) by
\begin{equation}
\label{eq:Gamma_LO_scalar}
    \Gamma_{\text{\tiny LO}}^{ \text{\tiny scalar}} = \frac{E_b}{\sqrt{2} \, \pi^2} \, \left(\frac{j_A r_0}{b} \right)^2 \, \sqrt{\frac{E}{E_b}} \, .
\end{equation}
where $j_A$ denotes the strength of the (non-conserved) source current for the scalar field.

In Figure~\ref{fig:Ga_s_Ga_p} we focus on the energy window \eqref{eq:GammaLO_zero_gap} where the decoherence rate vanishes (see also \eqref{eq:Gamma_zero_gap} for the generalization to higher orders) in the electromagnetic case (right panel), and we compare it with the scalar case (left panel), where such a window is absent. Note that the photon's curve exhibits a kink at the end of the coherence window, just like the absorption cross-section at the end of the transparency window \cite{Emparan:2025qqf}. This is a purely quantum gravity effect, since semiclassically the decoherence rate is always nonzero, at any value of the black hole energy above extremality $E$.

We also note that in both cases (the scalar and the photon) the decoherence rate above the transparency window ($E/E_b = 0$ for the scalar and $E/E_b = 1$ for the photon) behaves as $\sqrt{(E-E_{0,\ell})/E_b}$ for small positive $(E-E_{0,\ell})/E_b$.

\begin{figure}[t!]
    \centering
    \includegraphics[width=0.65\textwidth]{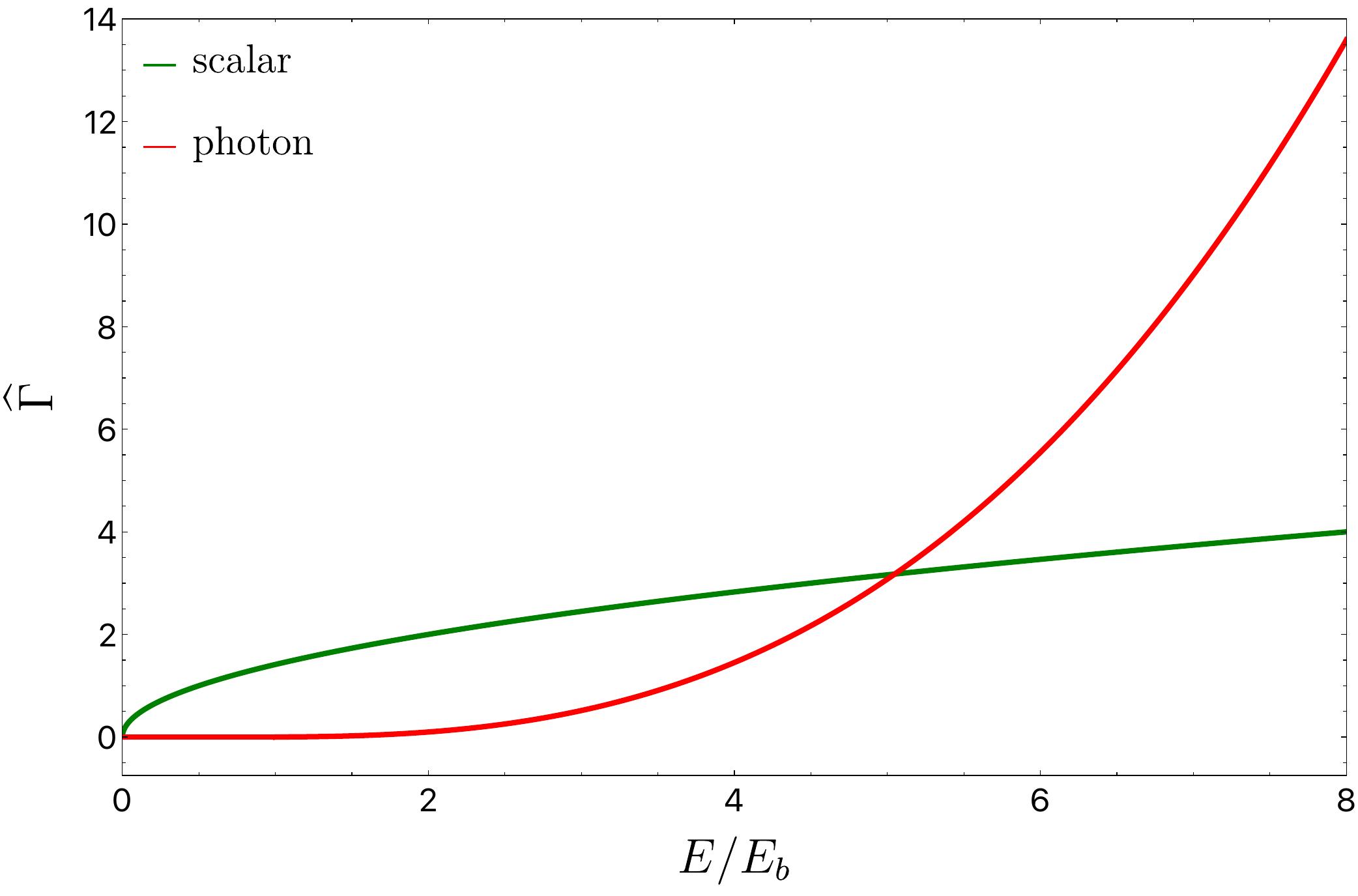}
    \caption{\small Schwarzian-corrected decoherence rate of the $\ell = 0$ scalar \eqref{eq:Gamma_LO_scalar} (green) vs.\ $\ell = 1$ photon \eqref{eq:GammaLO_fully_explicit} (red) as a function of $E/E_b$. Note that the electromagnetic decoherence rate surpasses the scalar one for $E \gtrsim 5 E_b$.}
    \label{fig:Ga_sp}
\end{figure}

In Figure~\ref{fig:Ga_sp} we compare the Schwarzian-corrected decoherence rates for the scalar and the photon. To make this comparison meaningful, we consider the quantity
\begin{equation}
\label{eq:Gamma_hat}
    \widehat{\Gamma} \equiv \frac{\Gamma}{2 \mathcal{N}^2 E_b^{2 \Delta - 1}} = E_b^{1-2 \Delta} \, \int_{-\infty}^{+\infty} dt \, \langle \mathcal{O}_\Delta^{(\ell,m)}(t) \mathcal{O}_\Delta^{(\ell,m)}(0) \rangle + \cdots \, ,
\end{equation}
which is independent of the type of field being considered, and at leading order is just given by the (dimensionless) $2$-point function of the response operators $\mathcal{O}_\Delta^{(\ell,m)}(t)$. So the competition between the scalar and the electromagnetic $\widehat{\Gamma}$ is only dictated by the interplay between angular momentum $\ell$ and conformal dimension $\Delta$ for the scalar ($\ell = 0$, $\Delta = 1$) and for the photon ($\ell = 1$, $\Delta = 3$) in the correlators of $\mathcal{O}_\Delta^{(\ell,m)}(t)$. In \eqref{eq:Gamma_hat}, $\mathcal{N}$ denotes the matching coefficient appropriate to the field type under consideration: $\mathcal{N}\propto e q d r_0^4/b^3$ for the photon, and $\mathcal{N}\propto j_A r_0/b$ for the scalar. As we see in Figure~\ref{fig:Ga_sp}, while for $E \lesssim E_b$ the scalar rate is larger, because of the photon's vanishing window \eqref{eq:GammaLO_zero_gap}, for $E \gtrsim 5 E_b$ the photon rate becomes the dominant one.

\subsection{Next-to-leading order, and conjecture for the higher orders}
\label{subsec:Gamma_NLO}

In the previous section we argued that the decoherence rate is zero for $E < E_b$ at leading (quadratic) order in $H_{\text{int}}$ due to the spin-induced gap \eqref{eq:gap} in the black hole spectrum. The next contribution to the decoherence rate in the expansion \eqref{eq:Phieqn2}, $\Gamma_{\text{\tiny NLO}}$, is quartic in $H_{\text{int}}$. If nonzero, this would be the leading contribution to the decoherence rate when $\Glo = 0$.

The computation of $\Gamma_{\text{\tiny NLO}}$ proceeds analogously to that of $\Gamma_{\text{\tiny LO}}$ and can be found in Appendix~\ref{app:rhoAT}. The result is
\begin{align}
    \Gamma_{\text{\tiny NLO}} \propto~& -\mathcal{N}^4 \, \sum_{j_3 = 0,1,2} \int_0^{+\infty} dE_2 \int_0^{+\infty} dE_3 \int_0^{+\infty} dE_4 \, \rho_1(E_2) \rho_{j_3}(E_3) \rho_1(E_4) \, \times\nonumber\\
    \label{eq:GammaNLO}
    &\left(\mathcal{O}_{12} \mathcal{O}_{23} \mathcal{O}_{34} \mathcal{O}_{41} \right)_c \, \left(2 \pi^2 \, \delta_{12} \delta_{13} \delta_{14} - \delta_{12} P_{13} P_{14} - \delta_{14} P_{12} P_{13} \right) \, ,
\end{align}
where we used the shorthand notation $\mathcal{O}_{n n'} \equiv \langle E_n, j_n, m_n^\text{\tiny BH} | \mathcal{O}(t=0) | E_{n'}, j_{n'}, m_{n'}^\text{\tiny BH} \rangle$, $\delta_{n n'} \equiv \delta(E_n - E_{n'})$, $P_{n n'} \equiv \text{PV}[1/(E_n - E_{n'})]$, and we have denoted $E_1 \equiv E$ to make the notation uniform. In particular, $\left(\mathcal{O}_{12} \mathcal{O}_{23} \mathcal{O}_{34} \mathcal{O}_{41} \right)_c$ denotes the connected $4$-point function of the response operators $\mathcal{O}(t) \equiv \mathcal{O}_3^{(1,0)}(t)$ in the microcanonical ensemble, whose explicit expression can be found for instance in \cite{Mertens:2017mtv,Jafferis:2022wez,Brown:2024ajk}.

Crucially, all the energy integrals in \eqref{eq:GammaNLO} contain either $\delta_{12}$ or $\delta_{14}$, or both, which have zero support when the black hole energy is inside the gap, $E \equiv E_1 < E_b$. This is because $j_2=j_4=1$, hence the densities $\rho_1(E_2) \propto \theta(E_2 - E_b)$ and $\rho_1(E_4) \propto \theta(E_4 - E_b)$ restrict the integration range to $\int_{E_b}^{+\infty} dE_2 \int_{E_b}^{+\infty} dE_4$. As a consequence, whenever $E < E_b$ the quantity $\Gamma_{\text{\tiny NLO}}$ vanishes identically, just like $\Gamma_{\text{\tiny LO}}$. Note that this argument is independent of the value of $j_3 = 0,1,2$. We conclude that the decoherence rate vanishes identically inside the gap up to quartic order in $H_{\text{int}}$:
\begin{equation}
\label{eq:GammaNLO_zero_gap}
    \Gamma_{\text{\tiny NLO}} = 0 \qquad \text{for } E < E_b \, .
\end{equation}

\paragraph{Connection to di-photon absorption.} 

The absorption rate at second order in perturbation theory involves four $\mathcal{O}$ insertions. One contribution to this rate comes from the absorption of two-photon spin-singlet states with total angular momentum zero (``di-photons''). The rate for di-photon absorption is nonzero even inside the gap because the black hole final state has $j = 0$. For $E < E_b$, it constitutes the leading contribution to the $\ell = 1$ photon absorption rate for a spinless black hole subject to unpolarized or linearly polarized radiation \cite{Brown:2024ajk,Emparan:2025qqf}.
One may then wonder if, similar to what happens at leading order (cf.~\eqref{eq:GammaLO_Gammaabs}), the next-to-leading order decoherence rate $\Gamma_{\text{\tiny NLO}}$ is controlled by the zero-frequency limit of the di-photon absorption rate, which would imply a nonzero contribution to the decoherence rate even inside the gap $E < E_b$.

However, this is not the case. As detailed in Appendix~\ref{app:rhoAT}, an intermediate contribution to $\Gamma_{\text{\tiny NLO}}$ has the form of a second-order transition rate,
\begin{equation}\la{trate}
    2\pi
    \left|
    \sum_I
    \frac{
    \langle E_3,0,0|\mathcal O|E_I,1,0\rangle
    \langle E_I,1,0|\mathcal O|E_1,0,0\rangle
    }{
    E_I-E_1
    }
    \right|^2
    \rho_0(E_3)
    \bigg|_{E_3\simeq E_1} ,
\end{equation}
which is precisely the structure expected for a two-photon virtual process.\footnote{To connect with the notation of \eqref{eq:GammaNLO}, the two integrals over $E_2$ and $E_4$ in \eqref{eq:GammaNLO} correspond to the two copies of the sum over intermediate state energies $E_I$ in \eqref{trate}.} However, this term does not survive as a contribution to the final expression of the decoherence rate $\Gamma_{\text{\tiny NLO}}$. In the full expression for $\Gamma_{\text{\tiny NLO}}$, the remaining terms generated by the influence functional expansion cancel this second-order transition rate contribution \eqref{trate}. As a consequence, the next-to-leading decoherence rate $\Gamma_{\text{\tiny NLO}}$ vanishes inside the gap, even though the di-photon absorption rate does not. Note that this cancellation occurs for any value of the black hole energy $E$, namely both below and above the gap, meaning that di-photon absorption never contributes to the decoherence rate.

One way to understand this result is as follows. The two orientations of Alice's dipole define two different Hamiltonians under which the initial state of the black hole evolves. If the black hole evolves into orthogonal states under these two possible Hamiltonian evolutions, decoherence occurs. These Hamiltonians are linear in the dipole  operator $P_B$, with a coefficient whose sign corresponds to the orientation of Alice’s dipole. Since when $E<E_{b}$ a two-photon process involves an intermediate off-shell state of the system, in the large $T$ limit, the time between subsequent $P_{B}$ insertions is forced to be very short. Therefore we effectively have an $\ell = 0$ composite operator containing two $P_{B}$ insertions.  Since this composite operator is quadratic in the dipole-dipole interaction, it is not sensitive to the sign of $P_A$. Therefore the Hamiltonian evolutions of the black hole state under the two branches of Alice's wavefunction are identical, and no decoherence occurs. In other words, the di-photon states with zero total angular momentum do not carry ``which path'' information about Alice's superposition that the black hole can collect by absorbing them.

\paragraph{Higher orders.}

More generally, while a black hole with $E<E_{b}$ can absorb spin-singlet multi-photon states, such processes will always involve an even number of $P_{B}$ insertions in the Dyson expansion of time evolution operator on each branch. More precisely, an odd number of $O(t)$ insertions can never create a state of total spin zero, since the symmetric tensor product of an odd number of spin 1 representations contains only states of odd total spin. Therefore, the black hole will again evolve identically under the two Hamiltonians defined by the two orientations of Alice's dipole. Based on this argument, we conjecture that the decoherence rate is zero at all orders in perturbation theory when $E < E_{b}$:
\begin{equation}
\label{eq:Gamma_zero_gap}
    \Gamma = 0 \qquad \text{for } E < E_b \, .
\end{equation}

\paragraph{Generality of the argument.} Our argument about the decoherence rate at all orders in perturbation theory applies to superpositions of a single charged particle. In this case, we can always think of Alice's wavefunction as the superposition of an electric dipole whose orientation differs by a $180^{\circ}$ spatial rotation across the two branches. However, it does not obviously prohibit decoherence of configurations involving more charges. For example, suppose Alice has access to two opposite sign charges and prepares a superposition of her dipole where the dipole orientation differs by an arbitrary (not $180^{\circ}$) rotation across the two branches. In this case, the Hamiltonian on each branch can differ by more than just a sign, and likewise spin-singlet multi-photon processes would be expected to cause decoherence even when $E<E_{b}$. Note that the decoherence rate at first perturbative order $\Gamma_{\text{\tiny LO}}$ in this case would still be zero below $E_b$ due to the spin-induced gap. Therefore, the value of the decoherence rate for this more general configuration would still be extremely suppressed relative to the semiclassical prediction.

\subsection{Comments on the relation to other observables}
\label{sec:rel_Green_abs}

To conclude this section, let us illustrate in more detail the relation between the decoherence rate and other observables controlled by the two-point function of $\mathcal{O}$. In this section, our interest is in comparing the $\omega$- and $E$-dependence of the various quantities, not their dimensions or normalizations. Likewise, the proportionality symbol $\propto$ here  denotes equality up to factors independent of the frequency $\omega$ and of the black hole energy $E$ above extremality. 

 As discussed in Section~\ref{sec:time_evol_RDM}, at leading order, the decoherence rate is proportional to the zero-frequency limit of the Wightman Green's function \cite{Biggs:2024dgp}:
\begin{equation}
\label{eq:Gamma_LO_sS}
    \Gamma_{\text{\tiny LO}} \propto \frac{e^2 q^2 d^2 r_0^8}{b^6} \, \mathcal{S}(\omega = 0)~~~~\text{ where }  ~~~~\mathcal{S}(\omega) = \int_{-\infty}^{+\infty} dt \, e^{i \omega t} \, \langle \mathcal{O}(t) \mathcal{O}(0) \rangle \, .
\end{equation}
The same Wightman function $\mathcal{S}(\omega)$ also controls the absorption and emission rates \cite{Emparan:2025sao,Biggs:2025nzs,Emparan:2025qqf} (cf.~\eqref{eq:GammaLO_Gammaabs}):
\begin{equation}
\label{eq:Gamma_abs_em_sS}
    \Gamma_\text{abs}(\omega) \propto \omega^3 \, \mathcal{S}(\omega) \, , \qquad \Gamma_\text{em}(\omega) \propto \omega^3 \, \mathcal{S}(-\omega) \, .
\end{equation}
On the other hand, from the optical theorem it follows that the absorption cross-section is determined by the combination
\begin{equation}
    \sigma_\text{abs}(\omega) \propto \frac{\Gamma_\text{abs}(\omega) - \Gamma_\text{em}(\omega)}{\omega^2} \propto \omega \, \big(\mathcal{S}(\omega) - \mathcal{S}(-\omega) \big)~~~~~~\omega r_0 \ll 1 \, .
\end{equation}
This combination corresponds to the imaginary part of the retarded Green's function
\begin{equation}
\label{eq:sGR_ImsGR}
    \mathcal{G}_R(\omega) = i \, \int_{-\infty}^{+\infty} dt \, e^{i \omega t} \, \theta(t) \, \langle [\mathcal{O}(t), \mathcal{O}(0)] \rangle \, , \qquad \text{Im}\,\mathcal{G}_R(\omega) = \frac{\mathcal{S}(\omega) - \mathcal{S}(-\omega)}{2} \, .
\end{equation}
The low-frequency behaviour of $\text{Im}\,\mathcal{G}_R(\omega)$ is captured, at first order in $\omega r_0$, by the dissipative response coefficients $\nu_\ell$ (also called ``dissipative Love numbers''):\footnote{In the context of near-extremal black branes in asymptotically AdS spacetime, the dissipative response coefficient associated with an $\ell = 0$ perturbation given by a minimally coupled massless neutral scalar field is related via Kubo's formula to the shear viscosity of the dual holographic fluid, whose quantum corrections have been studied recently \cite{Nian:2025oei,PandoZayas:2025snm,Cremonini:2025yqe,Gouteraux:2025exs,Kanargias:2025vul}.}
\begin{equation}
    \nu_1 \propto \lim_{\omega \to 0} \frac{\text{Im}\,\mathcal{G}_R(\omega)}{\omega} \propto \lim_{\omega \to 0} \frac{\sigma_\text{abs}(\omega)}{\omega^2} \propto \mathcal{S}'(\omega = 0) \, .
\end{equation}
Hence, in general the leading-order decoherence rate $\Gamma_{\text{\tiny LO}}$ is not controlled by the low-frequency dissipative response (or absorption cross-section), because $\mathcal{S}(0)$ and $\mathcal{S}'(0)$ are two independent quantities for a generic black hole state.

Nonetheless, for energies inside the gap $E < E_b$ we have $\mathcal{S}(-\omega) = 0$ \cite{Brown:2024ajk,Emparan:2025qqf}, so the emission rate $\Gamma_\text{em}(\omega)$ vanishes identically and we have $\sigma_\text{abs}(\omega) \propto \omega \, \mathcal{S}(\omega)$. Since $\mathcal{S}(\omega)$ vanishes when $E + \omega < E_b$ \cite{Emparan:2025qqf}, in the $\omega \to 0$ limit we have that both $\sigma_\text{abs}(\omega)$ and $\Gamma_{\text{\tiny LO}}$ vanish precisely when $E < E_b$. This is why the condition \eqref{eq:GammaLO_zero_gap} is just the zero-frequency limit of the transparency window discussed in \cite{Emparan:2025qqf} for the absorption cross-section, specialized to the $\ell = 1$ case.

An important case where $\Glo$ is fixed by the low-frequency behaviour of the absorption cross-section $\sigma_{\text{abs}}$ is when the black hole is in a thermal state, so the correlators satisfy the KMS condition. For fixed-energy states, this occurs in the semiclassical regime, where the microcanonical and canonical ensembles are equivalent. In this case, the KMS condition implies $\mathcal{S}_\text{sc}(-\omega) = e^{-\beta \omega} \, \mathcal{S}_\text{sc}(\omega)$, so we have
\begin{align}
    \sigma_{\text{abs}}(\omega) &\propto \omega \, \big(1-e^{-\beta \omega} \big) \, \mathcal{S}_\text{sc}(\omega) \approx \beta \omega^2 \, \mathcal{S}_\text{sc}(\omega)~~~~~~~ ~~~ \omega \beta \ll 1, ~~~~ \beta E_b \ll 1 \, ,
\end{align}
which is the fluctuation-dissipation theorem. As a consequence, in the semiclassical limit the leading-order decoherence rate is determined by the same dissipative response coefficient that controls the absorption cross-section at low frequency:
\begin{align}\la{absrel}
    \Glo \,\propto\, \frac{\nu_1}{\beta} \,\propto\, \lim_{\omega \to 0}\frac{\sigma_{\text{abs}}(\omega)}{\beta \omega^2}~~~~~~~ \beta \omega \ll 1, ~~~~ \beta E_b \ll 1 \, ,
\end{align}
as was already noted in \cite{Biggs:2024dgp}. However, this fluctuation-dissipation relation \eqref{absrel} does not generically hold at the quantum level (unless when working in the canonical ensemble \cite{Daguerre:2023cyx,Kanargias:2025vul}) or at higher orders in perturbation theory. Indeed, as discussed in the previous section, the decoherence rate and the low-frequency absorption cross-section behave differently at next order in perturbation theory.

\section{Closing remarks}
\label{sec:conclusions}

Previous work has demonstrated that non-extremal black holes decohere quantum systems in their exterior at a constant rate.  We showed that this is not a universal feature of black holes in quantum gravity.

The decoherence effect, which involves the absorption of soft radiation, relies on the presence of extremely fine level spacings. In semiclassical gravity, energy differences are so small, of the order of $\Delta E \sim e^{-S_0}$, that the density of states appears continuous to soft photons of frequencies larger than this tiny gap. However, at sufficiently low energies above extremality, the quantization of the black hole spin becomes relevant, and the density of states is sensitive to the much larger gaps between neighboring spin sectors, which scale as $\Delta E \sim 1/S_0^{\#}$. As a consequence, black holes sufficiently close to extremality become transparent to single soft photons, and do not decohere Alice's dipole. This is illustrated in the right panel of Figure~\ref{fig:Ga_s_Ga_p}. Of course, there also exist ordinary quantum systems with this property, such as an atom in a s-wave ground state with an energy gap to the first p-orbital excitation.

Note that a black hole transparent to single photons can still absorb two-photon states with total spin $0$. Curiously, this does not contribute to the decoherence. This is because such spin-singlet states do not contain ``which path'' information about the direction of Alice's dipole. In this respect, as an observable sensitive to the quantum gravity effects of near-extremal black holes, the decoherence is distinct from the absorption cross-section.

For energies $E$ above the gap, $E > E_b$, the decoherence rate no longer vanishes, but it remains smaller than the semiclassical prediction. This means that quantum gravity effects enhance the coherence of Alice's qubit both above and below the gap, as shown in Figure~\ref{fig:Ga_p_qusc}.

Here we analyzed decoherence mediated by the electromagnetic field. It would be interesting to consider the gravitational case where Alice's superposition is composed of two charge-neutral masses, and the effective coupling is between the mass quadrupole moments of the black hole and Alice's system. In this case we also expect an energy window where the leading order decoherence rate vanishes due to a spin-induced gap in the black hole spectrum. However, this analysis is complicated slightly by the mixing between $\ell \geq 2$ photon and graviton modes in the Reissner-Nordström background.

We also expect similar conclusions to hold for near-BPS black holes. For example, let us consider, as in \cite{Lin:2025wof}, a 4D asymptotically flat near-BPS black hole in ungauged $\mathcal{N} = 2$ supergravity. The black hole density of states contains a highly degenerate BPS sector at extremality ($E= 0$), which is separated by a gap from a continuum of near-BPS states organized into $\mathcal{N} = 4$ supermultiplets. 

In particular, the spins $j$ of the states in each supermultiplet are $J \oplus 2 (J-\frac{1}{2}) \oplus (J-1)$ where $J$ denotes the spin of the highest-spin state in the supermultiplet, and there is a gap $J^2 E_b / 2$ between the BPS zero-energy ground states (for which $E=0$ and $j=J=0$) and the first excited states belonging to the supermultiplet under consideration. Therefore, the first non-BPS states appear at $E = E_b/8$, in the $J = 1/2$ supermultiplet. Note that this is a short multiplet with one fewer state, containing spins $\frac{1}{2} \oplus 2(0)$. For a near-BPS spinless $(j=0)$ black hole in the $J = 1/2$ supermultiplet, in the photon channel we expect to have zero decoherence rate for $E_b/8 < E < E_b/2$. This is because single photon absorption would make the black hole transition to a $j = 1$ state, and the lowest energy $j = 1$ state lies in the $J = 1$ supermultiplet with energy $E>E_b/2$.

\section*{Acknowledgments}

We would like to thank Roberto Emparan,  Juan Maldacena, and Gautam Satishchandran for useful discussions and comments on the draft. The project that gave rise to these results received the support of a fellowship from ``la Caixa” Foundation (ID 100010434), awarded to ST, with code LCF/BQ/DI24/12070002.

\appendix

\section{Decoherence rate at fourth order in $H_\text{int}$}
\label{app:rhoAT}

In this appendix we will determine the explicit expression for the contribution at fourth order in $H_\text{int}$ to the influence functional $F(T)$ at late times. To this end, let us introduce the shorthand notation $\int dt_{1,\ldots,n} \equiv \int_0^T dt_1 \cdots \int_0^T dt_n$ and $\theta_{ij} \equiv \theta(t_i - t_j)$, where $\theta(x)$ denotes the Heaviside step function. Expanding the (anti-)time-ordered exponentials in \eqref{eq:FTdef} yields the following expansion of the influence functional (cf.~\eqref{eq:F0_F2}--\eqref{eq:F4_2}):
\begin{equation}
\label{eq:FT_expansion}
    F(T) = 1 + F_2(T) + F_4(T) + \cdots \, ,
\end{equation}
where
\begin{align}
    \label{eq:F2_time}
    F_2(T) &= - 2 \, \int dt_{1,2} \, \langle O(t_1) O(t_2) \rangle \, ,\\
    \label{eq:F4_time}
    F_4(T) &= \int dt_{1,2,3,4} \, \langle O(t_1) O(t_2) O(t_3) O(t_4) \rangle \, \Theta(t_1,t_2,t_3,t_4) \, .
\end{align}
Here we have defined the following combination of step functions:
\begin{align}
    \label{eq:Theta_1}
    \Theta(t_1,t_2,t_3,t_4) &= \theta_{12} \theta_{23} \theta_{34} + \theta_{43} \theta_{32} \theta_{21} + \theta_{21} \theta_{34} + \theta_{23} \theta_{34} + \theta_{32} \theta_{21}\\
    \label{eq:Theta_2}
    &= 2 \, \left(\theta_{12} \theta_{23} \theta_{34} + \theta_{43} \theta_{32} \theta_{21} + \theta_{21} \theta_{34} \right) \, ,
\end{align}
where the simplification occurs because of the property $\theta_{ji} = 1 - \theta_{ij}$.

As in the main text, we work in the microcanonical ensemble, but now we focus on the subleading contribution \eqref{eq:F4_time} (the leading contribution \eqref{eq:F2_time} is computed in Section~\ref{sec:Gamma_Sch}). To this end, let us denote the black hole state by $|E_1, j_1, m_1^\text{\tiny BH} \rangle$, with energy $E_1$, spin $j_1 = 0$ and azimuthal mode number $m_1^\text{\tiny BH} = 0$. To denote arbitrary intermediate states, let us introduce the shorthand notation $|n \rangle \equiv |E_n, j_n, m_n^\text{\tiny BH} \rangle$ and $\int dE_{2,\ldots,n} \equiv \int_0^{+\infty} dE_2 \cdots \int_0^{+\infty} dE_n$, where each energy integration must be weighted by the corresponding density of states $\rho_n \equiv \rho_{j_n}(E_n)$ (cf.~\eqref{eq:rho_j_E}). Inserting resolutions of the identity $\mathbb{I} = \int dE_n \, \rho_n \, |n \rangle \langle n |$, we can rewrite \eqref{eq:F4_time} in the energy basis as
\begin{equation}
\label{eq:F4_energy}
    F_4(T) = \int dt_{1,2,3,4} \int dE_{2,3,4} \, \rho_2 \rho_3 \rho_4 \, \langle 1| O(t_1) |2 \rangle \langle 2| O(t_2) |3 \rangle \langle 3| O(t_3) |4 \rangle \langle 4| O(t_4) |1 \rangle \, \Theta(t_1,t_2,t_3,t_4) \, .
\end{equation}
Introducing the notation $O_{n n'} \equiv \langle n | O(t=0) | n' \rangle$, the above expression simplifies to
\begin{equation}
\label{eq:F4_energy_2}
    F_4(T) = 2 \, \int dE_{2,3,4} \, \rho_2 \rho_3 \rho_4 \, O_{12} O_{23} O_{34} O_{41} \, I(T;E_1,E_2,E_3,E_4) \, ,
\end{equation}
where we have grouped the time integrals into the function
\begin{align}
    I(T;E_1,E_2,E_3,E_4) &= I_1 + I_2 + I_3 \, ,\\
    \label{eq:I1_def}
    I_1 &= \int dt_{1,2,3,4} \, e^{i (E_{12} t_1 + E_{23} t_2 + E_{34} t_3 + E_{41} t_4)} \, \theta_{12} \theta_{23} \theta_{34} \, ,\\
    \label{eq:I2_def}
    I_2 &= \int dt_{1,2,3,4} \, e^{i (E_{12} t_1 + E_{23} t_2 + E_{34} t_3 + E_{41} t_4)} \, \theta_{43} \theta_{32} \theta_{21} \, ,\\
    \label{eq:I3_def}
    I_3 &= \int dt_{1,2,3,4} \, e^{i (E_{12} t_1 + E_{23} t_2 + E_{34} t_3 + E_{41} t_4)} \, \theta_{21} \theta_{34} \, ,
\end{align}
where $E_{n n'} \equiv E_n - E_{n'}$. Note that $I_2 = (I_1)^\ast$, hence $I = 2 \, \text{Re}(I_1) + I_3$.

To extract the leading behaviour of the function $I$ for large $T$, it is convenient to consider the derivative $dI/dT$. In the large $T$ limit, we find the following expression for $dI_1/dT$:\footnote{Eq.~\eqref{eq:dI1_dT_1} can be obtained by performing the change of variables $(t_1,t_2,t_3,t_4) \mapsto (u_1,u_2,u_3,t_4)$, with $u_1 = t_1 - t_2$, $u_2 = t_2 - t_3$, $u_3 = t_3 - t_4$, in \eqref{eq:I1_def}.}
\begin{align}
    \label{eq:dI1_dT_1}
    \frac{dI_1}{dT} =~& \int_{\substack{u_1,u_2,u_3 > 0\\ u_1 + u_2 + u_3 < T}} du_1 du_2 du_3 \, e^{i (E_{12} u_1 + E_{13} u_2 + E_{14} u_3)}\\
    \label{eq:dI1_dT_2}
    \to~& \int_0^{+\infty} du_1 \int_0^{+\infty} du_2 \int_0^{+\infty} du_3 \, e^{i (E_{12} u_1 + E_{13} u_2 + E_{14} u_3)} \, .
\end{align}
Exploiting the distributional identity $\int_0^{+\infty} dt \, e^{i x t} = \pi \, \delta(x) + i \, \text{PV}\left(1/x \right)$, we find
\begin{equation}
    I_1 \sim T \, \left(\pi \delta_{12} + i P_{12} \right) \, \left(\pi \delta_{13} + i P_{13} \right) \, \left(\pi \delta_{14} + i P_{14} \right) + o(T) \, ,
\end{equation}
where we denoted $\delta_{n n'} \equiv \delta(E_n - E_{n'})$, $P_{n n'} \equiv \text{PV}[1/(E_n - E_{n'})]$. Here $o(T)$ denotes a term that is asymptotically smaller than $T$ as $T \to +\infty$, i.e. a function $f(T)$ such that $f(T)/T \to 0$ in that limit.

Similarly, for $I_3$ we find
\begin{equation}
    I_3 \sim 2 \pi T \, \delta_{13} \, \left(\pi \delta_{12} - i P_{12} \right) \, \left(\pi \delta_{14} + i P_{14} \right) + o(T) \, .
\end{equation}
Here we do not present in detail the computation of the large $T$ limit of $I_3$ because it is analogous to the one outlined above for $I_1$. The total time integral $I$ is then given, for large $T$, by
\begin{align}
    \label{eq:I_largeT_aux}
    I &= 2 \, \text{Re}(I_1) + I_3\\
    \label{eq:I_largeT}
    &\sim 2 \pi T \, \left[2 \pi^2 \, \delta_{12} \delta_{13} \delta_{14} - \delta_{12} P_{13} P_{14} - \delta_{14} P_{12} P_{13} + i \pi \delta_{13} \, \left(\delta_{12} P_{14} - \delta_{14} P_{12} \right) \right] + o(T) \, .
\end{align}
Note that the term $2 \pi T \delta_{13}P_{12}P_{14}$ appears both in $I_3$ and $2 ~\text{Re} (I_1)$ but with opposite sign. This term corresponds to the probability for di-photon absorption/emission  discussed in \eqref{trate}. While  such terms do appear in the calculation, they cancel in the sum \eqref{eq:I_largeT_aux}  and likewise do not in fact contribute to the influence functional.

At this point, it is important to note that, since our perturbation $\mathcal{O}$ carries $\ell = 1$ and $m=0$, composition of angular momenta yields $j_2 = 1$, $j_3 = 0,1,2$, $j_4 = 1$ and $m_2^\text{\tiny BH} = m_3^\text{\tiny BH} = m_4^\text{\tiny BH} = 0$. Plugging these values into \eqref{eq:F4_energy_2} and \eqref{eq:I_largeT}, we see that the contribution to $F(T)$ at fourth order in $H_\text{int}$ for large $T$ is given by
\begin{align}
    F_4(T) \sim~& 4 \pi T \, \sum_{j_3 = 0,1,2} \int dE_{2,3,4} \, \rho_1(E_2) \rho_{j_3}(E_3) \rho_1(E_4) \, O_{12} O_{23} O_{34} O_{41} \, \times\nonumber\\
    \label{eq:F4_largeT}
    &\left(2 \pi^2 \, \delta_{12} \delta_{13} \delta_{14} - \delta_{12} P_{13} P_{14} - \delta_{14} P_{12} P_{13} \right) + o(T) \, ,
\end{align}
where we noted that the two contributions to the imaginary part of $I$ in \eqref{eq:I_largeT} give rise to two equal and opposite integrals.

\section{Matching to the semiclassical decoherence rate}
\label{matching}

In this appendix we compute the semiclassical decoherence rate at leading order in our setup, and then we match it with the semiclassical limit of the quantum leading-order decoherence rate that we computed in Section~\ref{sec:Gamma_Sch}.

\paragraph{Semiclassical leading-order decoherence rate.} The relation \eqref{eq:Odef} between $O(t)$ and $\vec{P}_B(t)$ takes the following simple form in the particular configuration described in Section~\ref{sec:Gamma_Sch}:
\begin{equation}
    O(t) = \frac{e^2 q d}{4 \pi b^3} \, P_B^z(t) \, ,
\end{equation}
hence \eqref{eq:GammaLO} can be rewritten as \cite{Biggs:2024dgp}
\begin{equation}
\label{eq:GammaLO_PB}
    \Gamma_{\text{\tiny LO}} = 2 \, \left(\frac{e^2 q d}{4 \pi b^3} \right)^2 \, \lim_{\omega \to 0} \int_{-\infty}^{+\infty} dt \, e^{i \omega t} \, \langle P_B^z(t) P_B^z(0) \rangle \, .
\end{equation}
To proceed, let us define the retarded Green's function of the dipole operator $P_B^z(t)$:
\begin{equation}
\label{eq:chi_def}
    \chi(\omega) = i \, \int_{-\infty}^{+\infty} dt \, e^{i \omega t} \,  \theta(t) \, \langle [P_B^z(t), P_B^z(0)] \rangle \, .
\end{equation}

Up to this point our discussion was general, but let us now specialize to the semiclassical regime. In this regime, the microcanonical and canonical ensembles are equivalent, hence we can exploit the fluctuation-dissipation theorem \cite{Biggs:2024dgp}
\begin{equation}
\label{eq:fluct_diss}
    \text{Im}\,\chi_\text{sc}(\omega) = \frac{1 - e^{-\beta \omega}}{2} \, \int_{-\infty}^{+\infty} dt \, e^{i \omega t} \, \langle P_B^z(t) P_B^z(0) \rangle_\text{sc} \, ,
\end{equation}
where the subscript ``sc'' highlights that we are referring to the semiclassical correlators. Combining \eqref{eq:GammaLO_PB} with \eqref{eq:fluct_diss}, we find
\begin{equation}
\label{eq:Gamma_LO_Imchi_sc}
    \Gamma_{\text{\tiny LO}}^\text{sc} = \frac{4}{\beta} \, \left(\frac{e^2 q d}{4 \pi b^3} \right)^2 \, \lim_{\omega \to 0} \frac{\text{Im}\,\chi_\text{sc}(\omega)}{\omega} \, .
\end{equation}

On the other hand, the absorption cross-section is related as follows to the imaginary part of the retarded Green's function \cite{Goldberger:2005cd}: 
\begin{equation}
\label{eq:sigmaabs_Imchi_sc}
    \sigma_\text{abs}^\text{sc}(\omega) = e^2 \, \omega \, \text{Im}\,\chi_\text{sc}(\omega) \, ,
\end{equation}
so that combining \eqref{eq:Gamma_LO_Imchi_sc} with \eqref{eq:sigmaabs_Imchi_sc} we find
\begin{equation}
\label{eq:Gamma_LO_sigmaabs_sc}
    \Gamma_{\text{\tiny LO}}^\text{sc} = \frac{1}{\beta} \, \left(\frac{e q d}{2 \pi b^3} \right)^2 \, \lim_{\omega \to 0} \frac{\sigma_\text{abs}^\text{sc}(\omega)}{\omega^2} \, .
\end{equation}

The semiclassical absorption cross-section $\sigma_\text{abs}^\text{sc}(\omega)$ for the $\ell = 1$ photon is well-known in the literature \cite{Crispino:2000jx,Page:2000dk,Crispino:2009zza,Oliveira:2011zz,Arbey:2021jif, Arbey:2021yke}, and it is given by\footnote{Here we use the form presented in \cite{Emparan:2025qqf}, where $\sigma_\text{abs}^\text{ours}(\omega) = \sigma_\text{abs}^\text{theirs}(\omega)/3$.}
\begin{equation}
    \sigma_\text{abs}^\text{sc}(\omega) = \frac{4 \pi}{9} \, r_0^8 \, \omega^2 \, \left(\frac{4 \pi^2}{\beta^2} + \omega^2 \right) \, \left(\frac{16 \pi^2}{\beta^2} + \omega^2 \right) \, .
\end{equation}
As a consequence, from \eqref{eq:Gamma_LO_sigmaabs_sc} we obtain the explicit semiclassical expression for the leading-order decoherence rate:
\begin{equation}
\label{eq:Gamma_LO_sc_final_app}
    \Gamma_{\text{\tiny LO}}^\text{sc} = \frac{64 \pi^3}{9} \, \left(\frac{e q d r_0^4}{b^3} \right)^2 \, \beta^{-5} \, .
\end{equation}

\paragraph{Matching to the quantum expression.} In Section~\ref{subsec:Gamma_LO} we did not report the numerical prefactor in front of our expressions because that is specific to the particular configuration presented in Section~\ref{sec:Gamma_Sch}. Here we will report the precise prefactors because we want to match our quantum expression with the explicit semiclassical formula \eqref{eq:Gamma_LO_sc_final_app}.

By repeating the calculation described in Section~\ref{subsec:Gamma_LO}, but this time fixing the prefactors for our specific configuration, we obtain
\begin{align}
    \Gamma_{\text{\tiny LO}} &= 4 \pi \mathcal{N}^2 \, \rho_1(E) \, | \langle E, 0, 0 | \mathcal{O}(t=0) | E, 1, 0 \rangle |^2\\
    &= \mathcal{N}^2 \, \frac{\pi E_b^5}{120} \, \frac{\left(8 \, \frac{E}{E_b} + 1 \right)^2 \, \sinh\!\left(2 \pi \sqrt{\frac{2 (E - E_b)}{E_b}} \right)}{\cosh\!\left(2 \pi \sqrt{\frac{2 E}{E_b}} \right) - \cosh\!\left(2 \pi \sqrt{\frac{2 (E - E_b)}{E_b}} \right)} \, \theta(E - E_b) \, ,
\end{align}
whose semiclassical limit $E \gg E_b$ is given by
\begin{equation}
\label{eq:GammaLO_semicl_lim}
    \Gamma_{\text{\tiny LO}} \approx \mathcal{N}^2 \, \frac{32 \pi^5}{15} \, \beta^{-5} \, .
\end{equation}
Here we exploited the relation \eqref{eq:beta_E} between the energy and the inverse temperature of the black hole near extremality.

If we now compare \eqref{eq:Gamma_LO_sc_final_app} with the semiclassical limit \eqref{eq:GammaLO_semicl_lim} of our quantum expression, we find the value of the normalization constant $\mathcal{N}$:
\begin{equation}
\label{eq:sN_matching_value}
    \mathcal{N} = \frac{\sqrt{30}}{3 \pi} \, \frac{e q d r_0^4}{b^3} \, ,
\end{equation}
hence the quantum decoherence rate at leading order in our configuration has the explicit expression
\begin{equation}
\label{eq:GammaLO_fully_explicit}
    \Gamma_{\text{\tiny LO}} = \frac{E_b^5}{36 \pi} \, \left(\frac{e q d r_0^4}{b^3} \right)^2 \, \frac{\left(8 \, \frac{E}{E_b} + 1 \right)^2 \, \sinh\!\left(2 \pi \sqrt{\frac{2 (E - E_b)}{E_b}} \right)}{\cosh\!\left(2 \pi \sqrt{\frac{2 E}{E_b}} \right) - \cosh\!\left(2 \pi \sqrt{\frac{2 (E - E_b)}{E_b}} \right)} \, \theta(E - E_b) \, .
\end{equation}

\bibliography{Draft_v1/refs}

\providecommand{\href}[2]{#2}\begingroup\raggedright\begin{thebibliography}{10}

\bibitem{Biggs:2024dgp}
A.~Biggs and J.~Maldacena, \emph{{Comparing the decoherence effects due to black holes versus ordinary matter}},  \href{https://arxiv.org/abs/2405.02227}{{\ttfamily 2405.02227}}.

\bibitem{Danielson:2022tdw}
D.~L. Danielson, G.~Satishchandran and R.~M. Wald, \emph{{Black holes decohere quantum superpositions}}, \href{https://doi.org/10.1142/S0218271822410036}{\emph{Int. J. Mod. Phys. D} {\bfseries 31} (2022) 2241003} [\href{https://arxiv.org/abs/2205.06279}{{\ttfamily 2205.06279}}].

\bibitem{Danielson:2022sga}
D.~L. Danielson, G.~Satishchandran and R.~M. Wald, \emph{{Killing horizons decohere quantum superpositions}}, \href{https://doi.org/10.1103/PhysRevD.108.025007}{\emph{Phys. Rev. D} {\bfseries 108} (2023) 025007} [\href{https://arxiv.org/abs/2301.00026}{{\ttfamily 2301.00026}}].

\bibitem{Gralla:2023oya}
S.~E. Gralla and H.~Wei, \emph{{Decoherence from horizons: General formulation and rotating black holes}}, \href{https://doi.org/10.1103/PhysRevD.109.065031}{\emph{Phys. Rev. D} {\bfseries 109} (2024) 065031} [\href{https://arxiv.org/abs/2311.11461}{{\ttfamily 2311.11461}}].

\bibitem{Wilson-Gerow:2024ljx}
J.~Wilson-Gerow, A.~Dugad and Y.~Chen, \emph{{Decoherence by warm horizons}}, \href{https://doi.org/10.1103/PhysRevD.110.045002}{\emph{Phys. Rev. D} {\bfseries 110} (2024) 045002} [\href{https://arxiv.org/abs/2405.00804}{{\ttfamily 2405.00804}}].

\bibitem{Danielson:2024yru}
D.~L. Danielson, G.~Satishchandran and R.~M. Wald, \emph{{Local description of decoherence of quantum superpositions by black holes and other bodies}}, \href{https://doi.org/10.1103/PhysRevD.111.025014}{\emph{Phys. Rev. D} {\bfseries 111} (2025) 025014} [\href{https://arxiv.org/abs/2407.02567}{{\ttfamily 2407.02567}}].

\bibitem{Li:2024guo}
R.~Li, \emph{{Decoherence of quantum superpositions by Reissner-Nordstr{\"o}m black holes}}, \href{https://doi.org/10.1103/PhysRevD.111.024040}{\emph{Phys. Rev. D} {\bfseries 111} (2025) 024040} [\href{https://arxiv.org/abs/2411.04734}{{\ttfamily 2411.04734}}].

\bibitem{Li:2024lfv}
R.~Li, \emph{{Note on the local calculation of decoherence of quantum superpositions in de Sitter spacetime}}, \href{https://doi.org/10.1103/PhysRevD.111.044022}{\emph{Phys. Rev. D} {\bfseries 111} (2025) 044022} [\href{https://arxiv.org/abs/2501.00213}{{\ttfamily 2501.00213}}].

\bibitem{Danielson:2025iud}
D.~L. Danielson, J.~Kudler-Flam, G.~Satishchandran and R.~M. Wald, \emph{{How to minimize the decoherence caused by black holes}}, \href{https://doi.org/10.1103/67vv-km43}{\emph{Phys. Rev. D} {\bfseries 112} (2025) 025012} [\href{https://arxiv.org/abs/2501.04773}{{\ttfamily 2501.04773}}].

\bibitem{Kudler-Flam:2025yur}
J.~Kudler-Flam and G.~Penington, \emph{{It costs nothing to teleport information into a black hole}},  \href{https://arxiv.org/abs/2504.01058}{{\ttfamily 2504.01058}}.

\bibitem{Danielson:2025aji}
D.~L. Danielson and G.~Satishchandran, \emph{{Horizons and Soft Quantum Information}},  \href{https://arxiv.org/abs/2512.20754}{{\ttfamily 2512.20754}}.

\bibitem{Brown:2024ajk}
A.~R. Brown, L.~V. Iliesiu, G.~Penington and M.~Usatyuk, \emph{{The evaporation of charged black holes}},  \href{https://arxiv.org/abs/2411.03447}{{\ttfamily 2411.03447}}.

\bibitem{Mohan:2024rtn}
V.~Mohan and L.~Thorlacius, \emph{{Non-Perturbative Corrections to Charged Black Hole Evaporation}}, \href{https://doi.org/10.1007/JHEP04(2025)069}{\emph{JHEP} {\bfseries 04} (2025) 069} [\href{https://arxiv.org/abs/2411.13454}{{\ttfamily 2411.13454}}].

\bibitem{Maulik:2025hax}
S.~Maulik, X.~Meng and L.~A. Pando~Zayas, \emph{{Quantum-Corrected Hawking Radiation from Near-Extremal Kerr-Newman Black Holes}},  \href{https://arxiv.org/abs/2501.08252}{{\ttfamily 2501.08252}}.

\bibitem{Emparan:2025sao}
R.~Emparan, \emph{{Quantum cross-section of near-extremal black holes}}, \href{https://doi.org/10.1007/JHEP04(2025)122}{\emph{JHEP} {\bfseries 04} (2025) 122} [\href{https://arxiv.org/abs/2501.17470}{{\ttfamily 2501.17470}}].

\bibitem{Biggs:2025nzs}
A.~Biggs, \emph{{Following the state of an evaporating charged black hole into the quantum gravity regime}},  \href{https://arxiv.org/abs/2503.02051}{{\ttfamily 2503.02051}}.

\bibitem{Lin:2025wof}
G.~Lin, L.~V. Iliesiu and M.~Usatyuk, \emph{{The evaporation of black holes in supergravity}}, \href{https://doi.org/10.1007/JHEP08(2025)220}{\emph{JHEP} {\bfseries 08} (2025) 220} [\href{https://arxiv.org/abs/2504.21077}{{\ttfamily 2504.21077}}].

\bibitem{Emparan:2025qqf}
R.~Emparan and S.~Trezzi, \emph{{Quantum transparency of near-extremal black holes}}, \href{https://doi.org/10.1007/JHEP10(2025)023}{\emph{JHEP} {\bfseries 10} (2025) 023} [\href{https://arxiv.org/abs/2507.03398}{{\ttfamily 2507.03398}}].

\bibitem{Betzios:2025sct}
P.~Betzios, O.~Papadoulaki and Y.~Zhou, \emph{{Near-extremal quantum cross-section for charged fields and superradiance}},  \href{https://arxiv.org/abs/2507.13896}{{\ttfamily 2507.13896}}.

\bibitem{Kraus:2025efu}
P.~Kraus, \emph{{Hamiltonian approach to backreaction in near-extremal black hole evaporation}},  \href{https://arxiv.org/abs/2509.04293}{{\ttfamily 2509.04293}}.

\bibitem{Rakic:2025svg}
I.~Rakic, \emph{{The evaporation of near-extremal black holes through charged particle emission}},  \href{https://arxiv.org/abs/2512.18895}{{\ttfamily 2512.18895}}.

\bibitem{Luo:2026epp}
S.~Luo and L.~A. Pando~Zayas, \emph{{Quantum-Corrected Evaporation and Absorption Cross-Section of Near-Extremal Rotating Black Holes}},  \href{https://arxiv.org/abs/2601.06720}{{\ttfamily 2601.06720}}.

\bibitem{Satishchandran:2025cfk}
G.~Satishchandran, \emph{{Black Holes, Entanglement and Decoherence}},  in \emph{{24th International Conference on General Relativity and Gravitation (GR24) and 16th Edoardo Amaldi Conference on Gravitational (Amaldi16) Waves}}, 8, 2025, \href{https://arxiv.org/abs/2508.20171}{{\ttfamily 2508.20171}}.

\bibitem{Iliesiu:2020qvm}
L.~V. Iliesiu and G.~J. Turiaci, \emph{{The statistical mechanics of near-extremal black holes}}, \href{https://doi.org/10.1007/JHEP05(2021)145}{\emph{JHEP} {\bfseries 05} (2021) 145} [\href{https://arxiv.org/abs/2003.02860}{{\ttfamily 2003.02860}}].

\bibitem{Moitra:2019bub}
U.~Moitra, S.~K. Sake, S.~P. Trivedi and V.~Vishal, \emph{{Jackiw-Teitelboim Gravity and Rotating Black Holes}}, \href{https://doi.org/10.1007/JHEP11(2019)047}{\emph{JHEP} {\bfseries 11} (2019) 047} [\href{https://arxiv.org/abs/1905.10378}{{\ttfamily 1905.10378}}].

\bibitem{Iliesiu:2022onk}
L.~V. Iliesiu, S.~Murthy and G.~J. Turiaci, \emph{{Revisiting the logarithmic corrections to the black hole entropy}}, \href{https://doi.org/10.1007/JHEP07(2025)058}{\emph{JHEP} {\bfseries 07} (2025) 058} [\href{https://arxiv.org/abs/2209.13608}{{\ttfamily 2209.13608}}].

\bibitem{Kapec:2023ruw}
D.~Kapec, A.~Sheta, A.~Strominger and C.~Toldo, \emph{{Logarithmic Corrections to Kerr Thermodynamics}}, \href{https://doi.org/10.1103/PhysRevLett.133.021601}{\emph{Phys. Rev. Lett.} {\bfseries 133} (2024) 021601} [\href{https://arxiv.org/abs/2310.00848}{{\ttfamily 2310.00848}}].

\bibitem{Rakic:2023vhv}
I.~Rakic, M.~Rangamani and G.~J. Turiaci, \emph{{Thermodynamics of the near-extremal Kerr spacetime}}, \href{https://doi.org/10.1007/JHEP06(2024)011}{\emph{JHEP} {\bfseries 06} (2024) 011} [\href{https://arxiv.org/abs/2310.04532}{{\ttfamily 2310.04532}}].

\bibitem{Li:2025vcm}
R.~Li, Z.-X. Man and J.~Wang, \emph{{Decoherence of quantum superpositions in near-extremal Reissner-Nordstr{\"o}m black holes with quantum gravity corrections}}, \href{https://doi.org/10.1007/JHEP08(2025)079}{\emph{JHEP} {\bfseries 08} (2025) 079} [\href{https://arxiv.org/abs/2505.07480}{{\ttfamily 2505.07480}}].

\bibitem{Crispino:2009zza}
L.~C.~B. Crispino, A.~Higuchi and E.~S. Oliveira, \emph{{Electromagnetic absorption cross section of Reissner-Nordstrom black holes revisited}}, \href{https://doi.org/10.1103/PhysRevD.80.104026}{\emph{Phys. Rev. D} {\bfseries 80} (2009) 104026}.

\bibitem{Oliveira:2011zz}
E.~S. Oliveira, L.~C.~B. Crispino and A.~Higuchi, \emph{{Equality between gravitational and electromagnetic absorption cross sections of extreme Reissner-Nordstrom black holes}}, \href{https://doi.org/10.1103/PhysRevD.84.084048}{\emph{Phys. Rev. D} {\bfseries 84} (2011) 084048}.

\bibitem{Breuer}
H.~P. Breuer and F.~Petruccione, \emph{The Theory of Open Quantum Systems}. Oxford University Press, 01, 2007.

\bibitem{Bai:2023hpd}
Y.~Bai and M.~Korwar, \emph{{Near-extremal charged black holes: greybody factors and evolution}}, \href{https://doi.org/10.1007/JHEP03(2023)151}{\emph{JHEP} {\bfseries 03} (2023) 151} [\href{https://arxiv.org/abs/2301.07739}{{\ttfamily 2301.07739}}].

\bibitem{Mertens:2022irh}
T.~G. Mertens and G.~J. Turiaci, \emph{{Solvable models of quantum black holes: a review on Jackiw{\textendash}Teitelboim gravity}}, \href{https://doi.org/10.1007/s41114-023-00046-1}{\emph{Living Rev. Rel.} {\bfseries 26} (2023) 4} [\href{https://arxiv.org/abs/2210.10846}{{\ttfamily 2210.10846}}].

\bibitem{Heydeman:2020hhw}
M.~Heydeman, L.~V. Iliesiu, G.~J. Turiaci and W.~Zhao, \emph{{The statistical mechanics of near-BPS black holes}}, \href{https://doi.org/10.1088/1751-8121/ac3be9}{\emph{J. Phys. A} {\bfseries 55} (2022) 014004} [\href{https://arxiv.org/abs/2011.01953}{{\ttfamily 2011.01953}}].

\bibitem{Mertens:2017mtv}
T.~G. Mertens, G.~J. Turiaci and H.~L. Verlinde, \emph{{Solving the Schwarzian via the Conformal Bootstrap}}, \href{https://doi.org/10.1007/JHEP08(2017)136}{\emph{JHEP} {\bfseries 08} (2017) 136} [\href{https://arxiv.org/abs/1705.08408}{{\ttfamily 1705.08408}}].

\bibitem{Jafferis:2022wez}
D.~L. Jafferis, D.~K. Kolchmeyer, B.~Mukhametzhanov and J.~Sonner, \emph{{Jackiw-Teitelboim gravity with matter, generalized eigenstate thermalization hypothesis, and random matrices}}, \href{https://doi.org/10.1103/PhysRevD.108.066015}{\emph{Phys. Rev. D} {\bfseries 108} (2023) 066015} [\href{https://arxiv.org/abs/2209.02131}{{\ttfamily 2209.02131}}].

\bibitem{Nian:2025oei}
J.~Nian, L.~A. Pando~Zayas and C.-Y. Yue, \emph{{Quantum Corrections in the Low-Temperature Fluid/Gravity Correspondence}},  \href{https://arxiv.org/abs/2510.15411}{{\ttfamily 2510.15411}}.

\bibitem{PandoZayas:2025snm}
L.~A. Pando~Zayas and J.~Zhang, \emph{{One-loop Corrected Holographic Shear Viscosity to Entropy Density Ratio at Low Temperatures}},  \href{https://arxiv.org/abs/2510.16100}{{\ttfamily 2510.16100}}.

\bibitem{Cremonini:2025yqe}
S.~Cremonini, L.~Li, X.-L. Liu and J.~Nian, \emph{{Quantum Corrections to $\eta/s$ from JT Gravity}},  \href{https://arxiv.org/abs/2510.21602}{{\ttfamily 2510.21602}}.

\bibitem{Gouteraux:2025exs}
B.~Gout{\'e}raux, D.~M. Ramirez and C.~Supiot, \emph{{Schwarzian quantum corrections to shear correlators of the near-extremal Reissner-Nordstr{\"o}m-AdS black hole}},  \href{https://arxiv.org/abs/2512.19642}{{\ttfamily 2512.19642}}.

\bibitem{Kanargias:2025vul}
A.~Kanargias, E.~Kiritsis, S.~Murthy, O.~Papadoulaki and A.~P. Porfyriadis, \emph{{Holographic shear correlators at low temperatures, and quantum $\eta/s$}},  \href{https://arxiv.org/abs/2512.20443}{{\ttfamily 2512.20443}}.

\bibitem{Daguerre:2023cyx}
L.~Daguerre, \emph{{Boundary correlators and the Schwarzian mode}}, \href{https://doi.org/10.1007/JHEP01(2024)118}{\emph{JHEP} {\bfseries 01} (2024) 118} [\href{https://arxiv.org/abs/2310.19885}{{\ttfamily 2310.19885}}].

\bibitem{Goldberger:2005cd}
W.~D. Goldberger and I.~Z. Rothstein, \emph{{Dissipative effects in the worldline approach to black hole dynamics}}, \href{https://doi.org/10.1103/PhysRevD.73.104030}{\emph{Phys. Rev. D} {\bfseries 73} (2006) 104030} [\href{https://arxiv.org/abs/hep-th/0511133}{{\ttfamily hep-th/0511133}}].

\bibitem{Crispino:2000jx}
L.~C.~B. Crispino, A.~Higuchi and G.~E.~A. Matsas, \emph{{Quantization of the electromagnetic field outside static black holes and its application to low-energy phenomena}}, \href{https://doi.org/10.1103/PhysRevD.80.029906}{\emph{Phys. Rev. D} {\bfseries 63} (2001) 124008} [\href{https://arxiv.org/abs/gr-qc/0011070}{{\ttfamily gr-qc/0011070}}].

\bibitem{Page:2000dk}
D.~N. Page, \emph{{Thermodynamics of near extreme black holes}},  \href{https://arxiv.org/abs/hep-th/0012020}{{\ttfamily hep-th/0012020}}.

\bibitem{Arbey:2021jif}
A.~Arbey, J.~Auffinger, M.~Geiller, E.~R. Livine and F.~Sartini, \emph{{Hawking radiation by spherically-symmetric static black holes for all spins: Teukolsky equations and potentials}}, \href{https://doi.org/10.1103/PhysRevD.103.104010}{\emph{Phys. Rev. D} {\bfseries 103} (2021) 104010} [\href{https://arxiv.org/abs/2101.02951}{{\ttfamily 2101.02951}}].

\bibitem{Arbey:2021yke}
A.~Arbey, J.~Auffinger, M.~Geiller, E.~R. Livine and F.~Sartini, \emph{{Hawking radiation by spherically-symmetric static black holes for all spins. II. Numerical emission rates, analytical limits, and new constraints}}, \href{https://doi.org/10.1103/PhysRevD.104.084016}{\emph{Phys. Rev. D} {\bfseries 104} (2021) 084016} [\href{https://arxiv.org/abs/2107.03293}{{\ttfamily 2107.03293}}].

\end{thebibliography}\endgroup

\end{document}